\documentclass[aps,pre,reprint,superscriptaddress,amsmath,amssymb,floatfix,longbibliography]{revtex4-2}
\usepackage[T1]{fontenc}
\usepackage{graphicx}
\usepackage{bm}
\usepackage{mathtools}
\usepackage{xcolor}
\usepackage{hyperref}
\definecolor{darkblue}{rgb}{0,0,0.6}
\hypersetup{colorlinks,linkcolor=darkblue,citecolor=darkblue,urlcolor=darkblue}
\usepackage{mathrsfs}

\makeatletter
\newcommand{\lpmb}[1]{\mathpalette\lpmb@{#1}}
\newcommand{\lpmb@}[2]{%
  \ooalign{%
    $\m@th#1#2$\cr
    \hidewidth\kern0.16pt
    \raise0.08pt\hbox{$\m@th#1#2$}%
    \hidewidth\cr
  }%
}
\makeatother

\newif\ifgrayresultboxes
\grayresultboxestrue
\ifgrayresultboxes
  \usepackage{tcolorbox}
  \newcommand{\resultbox}[1]{%
    \begin{tcolorbox}[width=\linewidth,colback=gray!5,colframe=black!35,
      boxrule=0.5pt,arc=2mm,left=3pt,right=3pt,top=3pt,bottom=3pt]
      \centering #1
    \end{tcolorbox}%
  }
\else
  \newcommand{\resultbox}[1]{\par\medskip\noindent #1\par\medskip}
\fi
\newcommand{\T}{\mathsf T}
\newcommand{\mathv}{v}
\newcommand{\stiffness}{\bm{\mathcal K}}

\begin{document}
\title{Topology of Fluctuation Bands in Chiral Active Crystals}
\author{Rapha\"el Maire}
\email{maire@ub.edu}
\affiliation{Departament de F\'isica de la Mat\`eria Condensada,
Universitat de Barcelona, Mart\'i Franqu\`es 1, 08028 Barcelona, Spain}
\begin{abstract}
While band topology is usually associated with the deterministic dynamics of a system, we show that it can instead reside in its fluctuations. Using a reciprocal lattice driven by nonequilibrium chiral active noise, we find that the displacement spectrum---which quantifies displacement fluctuations---admits bands with nonzero Chern numbers despite a Chern-trivial deterministic mechanics. A two-band valley theory captures the emergence of this topology through a Dirac mass inversion, a Haldane-type mechanism, or a quadratic band touching, depending on the relevant transition. Through the bulk--boundary correspondence, we find boundary-localized fluctuation modes that are distinct from the usual propagating mechanical edge states.
\end{abstract}
\maketitle

\section{Introduction}
\label{sec:introduction}

Topological band theory classifies global properties of eigenspaces that remain unchanged as long as the relevant spectral gaps and protecting symmetries are preserved. This robustness allows for generic, nonperturbative predictions that remain valid in realistic, noisy systems. In quantum matter, Berry curvature and Chern numbers explain the quantized Hall response and, through the bulk--boundary correspondence, enforce special states at the edges of topological materials~\cite{berry1984quantal,simon1983holonomy,klitzing1980new,laughlin1981quantized,thouless1982quantized,niu1985quantized,hatsugai1993chern}. Their manifestations depend on the relevant spectral gap and protecting symmetries~\cite{haldane1988model,schnyder2008classification,ryu2010topological,chiu2016classification}. Although developed in quantum physics, topological band theory can apply to any system with a gapped eigenspace. Accordingly, photonic, acoustic, electrical, and mechanical systems can display nontrivial topological properties, including protected waveguiding, robust edge transport, boundary and corner states, defect-insensitive propagation, and robust zero modes~\cite{haldane2008possible,wang2009observation,rechtsman2013photonic,prodan2009topological,mousavi2015elastic,wang2015oneway,susstrunk2015phononic,kane2014topological,paulose2015modes,nash2015gyroscopic,rocklin2016mechanical,serragarcia2018observation,ningyuan2015time,imhof2018topolectrical,ghatak2020observation}.

Nonequilibrium classical systems are likewise good candidates for topological classification. Because detailed balance is broken~\cite{marconi2008fluctuation,seifert2012stochastic,gnesotto2018broken}, the evolution operator becomes a natural object of study. Its topological properties can govern pumping, transport, and localization in stochastic and dissipative systems~\cite{sinitsyn2009stochastic,murugan2017topologically,dasbiswas2018topological,tang2021topology}. Active matter provides a natural setting for such investigations, since its dynamics are driven by local energy consumption~\cite{marchetti2013hydrodynamics,bechinger2016active,cates2024active}. Because biological active materials are intrinsically noisy and heterogeneous~\cite{needleman2017active,gompper2020physics,aranson2022bacterial}, topology offers a robust framework for classifying their collective behavior. In such active matter systems, topology generally refers either to real-space defects or to bands of collective modes~\cite{bowick2022symmetry,shankar2022topological,sone2026hermitian}.

Real-space topological defects are already important to classical equilibrium systems, where they underlie phase transitions in two-dimensional systems with continuous degrees of freedom, including the BKT transition of the XY model and some two-dimensional melting~\cite{halperin1978melting,nelson1979dislocation,kosterlitz1973ordering,chaikin1995principles}. Their principal application in active matter is related to active nematics, in which continuously created and annihilated $\pm 1/2$ disclinations can order, generate active turbulence, and exhibit nonequilibrium dynamics such as defect self-propulsion~\cite{sanchez2012spontaneous,genkin2017topological,sokolov2019emergence,sokolov2025synthetic, giomi2013defect, PhysRevLett.113.038302,shankar2019hydrodynamics,keber2014topology}. These defects can also control biological functions~\cite{saw2017topological,duclos2020topological,yashunsky2024topological}. Topological defects are important in many other active systems as well~\cite{radhakrishnan2026irreversibility,casagrande2026topologypulsatingactivematter,banerjee2025contraction,rouzaire2026dynamics,rouzaire2025nonreciprocal,PhysRevLett.127.088004,PhysRevLett.134.188301,10.3389/fphy.2022.976515,rana2024defect,bililign2022motile,PhysRevLett.127.268001}.

The second use of topology in active matter concerns the band topology of operators governing collective dynamics, often after linearization. Topological bands and their boundary modes can occur in circulating liquids, curved flocks, and periodic active flows~\cite{souslov2017topological,shankar2017topological,sone2019anomalous,shankar2022topological,scheibner2020nonhermitian,uchida2026designing,coulais2021topology,rvkr-7mr2,sone2020exceptional}. Here the invariant belongs to a dynamical or transport operator. Depending on the dynamics, this operator may define either a Hermitian band problem, with a real spectrum and orthogonal eigenspaces, or a non-Hermitian one, with complex eigenvalues and distinct left and right eigenvectors.

Chiral active fluids provide a realization of such band topology. In \textit{some} linearized chiral hydrodynamic models, intrinsic rotation gaps density--velocity bands through a Coriolis-like coupling, while odd viscosity regularizes the short-wavelength response. Together, they allow a well-defined Chern number and the corresponding bulk--boundary correspondence~\cite{souslov2019topological,banerjee2017odd,ganeshan2017odd,tauber2020anomalous,fujii2025gauge}. An analogous structure appears in rotating shallow-water dynamics, where the equatorial Kelvin and Yanai waves are topological~\cite{delplace2017topological,perrot2019topological,tauber2019bulk,tauber2020anomalous,venaille2021wave}. Odd viscosity is not necessary for topological states in chiral active matter. For instance, chiral random walkers, including models with chirality switching, can display topological states~\cite{wojcik2026chiral,osat2026topological,kuroda2026designing,edwards2026robusttopologicallyprotectededge}.

These topological boundary waves should not, however, be conflated with the steady edge currents commonly observed in chiral active matter~\cite{caprini2019active,yang2020robust,beppu2021edge,yashunsky2022chiral,li2024robust,caporusso2024phase,langford2025phase,petrini2026curvaturedrivenwallaccumulationchiral}. Although the two phenomena are often discussed together~\cite{yamauchi2020chirality} and may coexist~\cite{kuroda2026designing,edwards2026robusttopologicallyprotectededge,poggioli2023emergent}, they often have distinct origins. Edge currents typically arise from a jump in torque density across an interface, possibly a wall~\cite{metzger2026equationstateedgeflow, Marconi2026hydrodynamics,maire2026kinetic2,reyes2026steady}. By contrast, a topological edge mode is a boundary-localized state enforced by a bulk invariant and need not carry a sustained material flux.

These examples attribute topology to the deterministic dynamics, with nonequilibrium chirality built directly into the time evolution operator. Active matter, however, displays strong nonequilibrium fluctuations---not constrained by a fluctuation-dissipation relation---leaving open a new possibility: \textit{topology carried by the fluctuations themselves}.
To make this idea concrete, consider a lattice and the lattice displacement field $\bm U(\bm k,t)$ in Fourier space with wavenumber $\bm k$. Its noisy time evolution, after linearization, is
\begin{equation}
\dot{\bm U}(\bm k,t)=\mathcal L(\bm k)\bm U(\bm k,t)+\bm\xi(\bm k,t).
\end{equation}
Conventionally, topology is assigned to the deterministic operator $\mathcal L$, while the noise $\bm\xi$ is regarded as merely exciting its modes. Here we instead treat the fluctuations themselves as a band structure. At fixed observation frequency $\omega$, these fluctuations are encoded by the displacement spectrum $\bm S(\bm k,\omega)$:
\begin{equation}
\bm S(\bm k,\omega)\propto \left\langle\bm U(\bm k,\omega)\bm U^\dagger(\bm k',\omega')\right\rangle.
\end{equation}
We ask whether the eigenvectors of $\bm S$ can acquire nontrivial Chern topology even when both the deterministic dynamics and the noise spectrum are individually Chern-trivial. In this article, we show that they can.
\resultbox{The displacement spectrum $\bm S(\bm k,\omega)$ can be topologically nontrivial, even when neither the deterministic dynamics $\mathcal L(\bm k)$ nor the noise $\bm \xi(\bm k, \omega)$ is topological.}
In Sec.~\ref{sec:model}, we study reciprocal elastic lattices driven by local chiral Ornstein--Uhlenbeck forces, so that the mechanical stiffness and deterministic generator are Chern-trivial and all handedness enters through the random noise. We find that chiral random forcing, combined with the reciprocal elastic response, produces correlation bands with nonzero Chern numbers. In the honeycomb lattice, their topological onset follows a Haldane-type Dirac mass inversion. The Chern number is tunable through the drive handedness, persistence, and observation frequency (Sec.~\ref{sec:topology}). A given single system can therefore have Chern-trivial fluctuations when observed at frequency $\omega_1$ but Chern-nontrivial fluctuations at frequency $\omega_2$. We then establish the corresponding bulk--boundary relation for boundary-localized fluctuations (Sec.~\ref{sec:bulk-boundary}) and show that the construction extends to various lattices, including three-dimensional ones (Sec.~\ref{sec:discussion}).

\section{Model}
\label{sec:model}

We now introduce a specific lattice and stochastic dynamics. Section~\ref{sec:discussion} later separates the ingredients required by the mechanism from those chosen only for analytical and numerical convenience.

\subsection{Real-space dynamics and notation}
\label{sec:real-space-model}

\begin{figure*}[t]
    \centering
    \includegraphics[width=\linewidth]{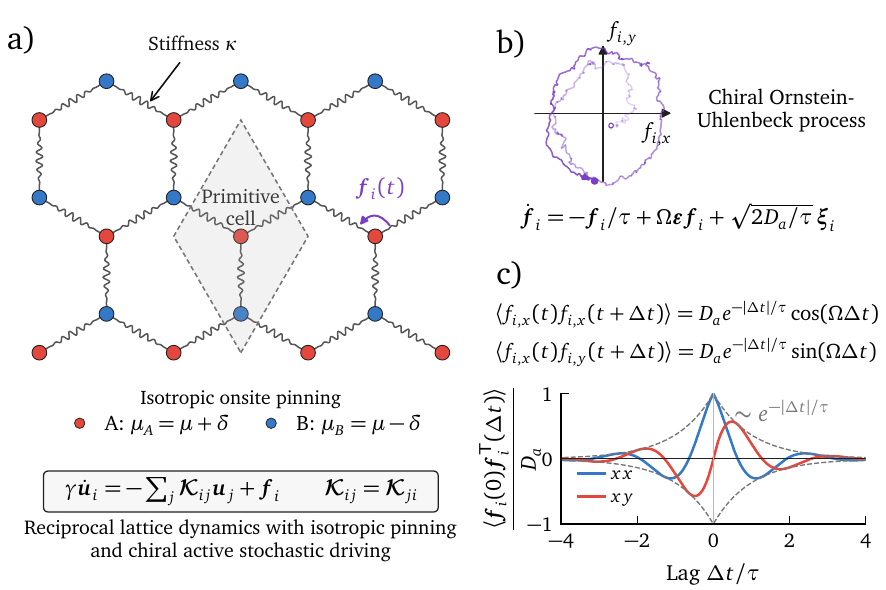}
    \caption{\textbf{Reciprocal lattice and rotating active forcing.} \textbf{a)} Honeycomb lattice with reciprocal nearest-neighbor springs driven by rotating active forces, with two sites $A$ and $B$ per primitive unit cell and sublattice-dependent pinning $\mu_A$ and $\mu_B$. \textbf{b)} A realization of the rotating active force $\bm f_i$, where $i=(n,\alpha)$ labels a lattice site. \textbf{c)} Longitudinal and transverse entries of the covariance $\bm q(\Delta t)$.}
    \label{fig:system-overview}
\end{figure*}

We consider a periodic two-dimensional honeycomb lattice whose primitive unit cells are indexed by $n$ and whose two sites per cell are indexed by $\alpha\in\{A, B\}$, as shown in Fig.~\ref{fig:system-overview}(a). Each site moves in the plane with displacement
\begin{equation}
\bm u_{n\alpha}(t)= \begin{pmatrix} u_{n\alpha,x}(t)\\
u_{n\alpha,y}(t) \end{pmatrix},
\end{equation}
and obeys the overdamped dynamics
\begin{equation}
\gamma\dot{\bm u}_{n\alpha}=-\sum_{m,\beta} \stiffness_{n\alpha,m\beta}\bm u_{m\beta} + \bm f_{n\alpha}.
\label{eq:real_space_dynamics}
\end{equation}
Nearest-neighbor sites are connected by \emph{central springs} of stiffness $\kappa$. For later convenience, the two sublattices are also pinned isotropically with the stiffnesses shown in Fig.~\ref{fig:system-overview}(a):
\begin{equation}
\mu_A=\mu+\delta,\qquad\mu_B=\mu-\delta,
\end{equation}
such that the elastic energy is
\begin{equation}
\begin{aligned}
E ={}& \frac{\kappa}{2} \sum_{\langle n\alpha,m\beta\rangle} \left[ \hat{\bm n}_{n\alpha,m\beta} \cdot \left(\bm u_{n\alpha}-\bm u_{m\beta}\right) \right]^2\\
&+ \frac{1}{2}\sum_n \left( \mu_A|\bm u_{nA}|^2 + \mu_B|\bm u_{nB}|^2 \right).
\end{aligned}
\end{equation}
Here the first sum runs over all nearest-neighbor bonds, and $\hat{\bm n}_{n\alpha,m\beta}$ is the unit vector along a bond. The real-space stiffness matrix $\stiffness$ in Eq.~\eqref{eq:real_space_dynamics} is the Hessian of this energy and contains all interactions. It is symmetric by construction and, on its own, is fully compatible with equilibrium dynamics.

To drive the system far from equilibrium, each site is subjected to an independent random force $\bm f$ that follows a chiral Ornstein--Uhlenbeck process~\cite{caprini2019active,caprini2023chiral,
sahala2025self,deion2026chiral}, illustrated in Fig.~\ref{fig:system-overview}(b),
\begin{equation}
\begin{gathered}
\tau\dot{\bm f}_{n\alpha} = -\bm f_{n\alpha} +\Omega\tau \bm\varepsilon\bm f_{n\alpha} +\sqrt{2D_a\tau} \bm\xi_{n\alpha}, \quad \bm\varepsilon= \begin{pmatrix} 0 & -1 \\
1 & 0 \end{pmatrix}\!.
\end{gathered}
\label{eq:ou}
\end{equation}
The white noise $\bm\xi_{n\alpha}$ driving the active force has zero mean and unit delta-covariance:
\begin{gather}
\langle\bm\xi_{n\alpha}(t)\rangle=0,\\
\langle\xi_{n\alpha,p}(t)\xi_{m\beta,q}(t')\rangle=\delta_{nm}\delta_{\alpha\beta}\delta_{pq}\delta(t-t'),\quad p,q\in\{x,y\}.\notag
\end{gather}
Here $\tau$ is the persistence time, and $\Omega$ is the angular rotation rate of the active force whose sign fixes the rotational handedness. Equivalently, $\bm f$ is a zero-mean, time-correlated Gaussian noise with the covariance shown in Fig.~\ref{fig:system-overview}(c):
\begin{equation}
\begin{gathered}
\left\langle \bm f_{n\alpha}(t) \bm f_{m\beta}^{\T}(t+\Delta t) \right\rangle = \delta_{nm}\delta_{\alpha\beta} \bm q(\Delta t),\\
\bm q(\Delta t) = D_ae^{-|\Delta t|/\tau} \begin{pmatrix} \cos(\Omega\Delta t)&\sin(\Omega\Delta t)\\
-\sin(\Omega\Delta t)&\cos(\Omega\Delta t) \end{pmatrix}.
\end{gathered}
\end{equation}
Because the elastic interactions derive from a global energy, the random drive $\bm f$ is the only source of nonequilibrium behavior. At finite persistence, this colored forcing is not paired with a corresponding memory kernel and therefore violates the fluctuation--dissipation theorem. A nonzero $\Omega$ additionally breaks time-reversal symmetry through the chirality of the drive. 

Our model therefore describes a crystal of chiral active particles. They are known to exhibit rich phenomena including long-range correlations, collective states, and instabilities~\cite{tan2022odd,shee2024emergent,kuroda2024long,marconi2025spontaneous,musacchio2026circling,caprini2025odd,caprini2024self}.

We now analyze the fluctuating properties of this crystal.

\subsection{Wavevectors, force spectrum, and displacement spectrum}
\label{sec:spectral-objects}

Taking the nearest-neighbor distance as the unit of length, we choose the primitive honeycomb Bravais vectors
\begin{equation}
\begin{gathered}
\bm a_1 = \begin{pmatrix}\sqrt3/2\\
3/2\end{pmatrix}, \qquad \bm a_2= \begin{pmatrix}-\sqrt3/2\\
3/2\end{pmatrix},
\end{gathered}
\end{equation}
with $\bm R_n=n_1\bm a_1+n_2\bm a_2$ the Bravais-lattice position of the cell $n$. The Bravais lattice constant used below is $a\equiv|\bm a_1|=|\bm a_2|=\sqrt3$ in these units. Their reciprocal vectors $\bm b_i$ are defined by $\bm a_i\cdot\bm b_j=2\pi\delta_{ij}$,
\begin{equation}
\bm b_1 = \begin{pmatrix}2\pi/\sqrt3\\
2\pi/3\end{pmatrix}, \qquad \bm b_2 = \begin{pmatrix}-2\pi/\sqrt3\\
2\pi/3\end{pmatrix}.
\end{equation}
A point in reciprocal space is decomposed by $\bm k=u\bm b_1+\mathv\bm b_2$, so that $\bm k\cdot\bm a_1=2\pi u$ and $\bm k\cdot\bm a_2=2\pi \mathv$. Useful points in the Brillouin zone used below are (see Fig.~\ref{fig:bands}(e))
\begin{equation}
\begin{gathered}
\bm{\Gamma}=\bm0,~ \bm{K}=\frac{\bm b_1+2\bm b_2}{3},~
\bm{M}=\frac{\bm b_1+\bm b_2}{2},~ \bm{K}'=\frac{2\bm b_1+\bm b_2}{3}.
\end{gathered}
\end{equation}

For a periodic sample with $N_{\rm c}$ unit cells, we define the Fourier transform by
\begin{equation}
\begin{aligned}
\bm u_\alpha(\bm k,\omega) &=\sum_n\int_{-\infty}^{\infty}dt\, \bm u_{n\alpha}(t)e^{\mathrm i\omega t-\mathrm i\bm k\cdot\bm R_n},\\
\bm u_{n\alpha}(t) &=\frac{1}{N_{\rm c}}\sum_{\bm k}\int_{-\infty}^{\infty}\frac{d\omega}{2\pi}\, \bm u_\alpha(\bm k,\omega)e^{-\mathrm i\omega t+\mathrm i\bm k\cdot\bm R_n}.
\end{aligned}
\label{eq:fourier_convention}
\end{equation}
The wavevector sum runs over the $N_{\rm c}$ points in the first Brillouin zone. We use the same transform for $\bm f_{n\alpha}$ and collect the two sublattice vectors into the four-component unit-cell fields
\begin{equation}
\bm U = \begin{pmatrix} \bm u_A\\
\bm u_B \end{pmatrix} = \begin{pmatrix} u_{Ax}\\
u_{Ay}\\
u_{Bx}\\
u_{By} \end{pmatrix}, \qquad \bm F= \begin{pmatrix} \bm f_A\\
\bm f_B \end{pmatrix}.
\end{equation}
The arguments $(\bm k,\omega)$ are suppressed when unambiguous, and Fourier-transformed quantities use the same symbols as their real-space counterparts.

Fourier transforming Eq.~\eqref{eq:real_space_dynamics} gives
\begin{equation}
\begin{gathered}
\bm U(\bm k,\omega) = \bm G(\bm k,\omega)\bm F(\bm k,\omega), \\
\bm G(\bm k,\omega) = \left[ \stiffness(\bm k)-\mathrm i\gamma\omega\bm I_4 \right]^{-1},
\end{gathered}
\label{eq:propagator}
\end{equation}
where $\bm G$ is the $4\times4$ propagator. The Bloch stiffness $\stiffness(\bm k)$ is Hermitian and positive semidefinite because it derives from the elastic energy~\footnote{It is positive definite only when pinning removes all zero modes.}. 

The Fourier transform $\bm q(\omega)=\int_{-\infty}^{\infty}\bm q(\Delta t)e^{-\mathrm i\omega\Delta t}d\Delta t$ of the single-site force autocorrelation is diagonal in the circular basis
\begin{equation}
\bm e_\pm=\frac{1}{\sqrt2} \begin{pmatrix}1\\
\pm\mathrm i\end{pmatrix}, \qquad f_\pm = \bm e_\pm^\dagger\bm f,
\label{eq:circular_basis}
\end{equation}
with
\begin{equation}
\begin{gathered}
\bm q(\omega)=q_+\bm e_+\bm e_+^\dagger+q_-\bm e_-\bm e_-^\dagger=q_0(\omega)\left[\bm I_2+\mathrm i\chi(\omega)\bm\varepsilon\right],\\
q_\pm(\omega)\equiv\bm e_\pm^\dagger\bm q(\omega)\bm e_\pm=\frac{2D_a\tau}{1+\tau^2(\omega\mp\Omega)^2}
\end{gathered}
\label{eq:q_def}
\end{equation}
where
\begin{equation}
\begin{gathered}
q_0=\frac{q_++q_-}{2},\quad \chi=\frac{q_+-q_-}{q_++q_-}=\frac{2\omega\Omega\tau^2}{1+\tau^2(\omega^2+\Omega^2)}.
\end{gathered}
\end{equation}
Here $q_0>0$ controls the driving intensity, while $\chi\in[-1,1]$ is the normalized circular-polarization imbalance of the force spectrum at observation frequency $\omega$. We henceforth call $\chi$ the \emph{spectral polarization}. Unlike the microscopic rotation rate $\Omega$, it is a bounded, frequency-dependent quantity determined by $\omega$, $\Omega$, and $\tau$. Because the two sublattices are forced independently, the full $4\times4$ active-force spectrum in the basis $(Ax,Ay,Bx,By)$ is
\begin{equation}
\bm Q(\omega)=\bm I_{\mathrm{AB}}\otimes\bm q(\omega) = \begin{pmatrix} \bm q(\omega)&\bm0_2\\
\bm0_2&\bm q(\omega) \end{pmatrix},
\label{eq:full_force_covariance}
\end{equation}
with  $\bm Q$ defined by the spectrum of the random driving:
\begin{equation}
\begin{gathered}
\left\langle \bm F(\bm k,\omega)\bm F^\dagger(\bm k',\omega')\right\rangle =2\pi N_{\rm c} \delta_{\bm k,\bm k'}\delta(\omega-\omega')\bm Q(\omega).
\end{gathered}
\label{eq:force_spectral_density}
\end{equation}
Using Eqs.~\eqref{eq:propagator} and \eqref{eq:force_spectral_density}, we similarly define the displacement  spectrum by
\begin{equation}
\begin{gathered}
\left\langle \bm U(\bm k,\omega)\bm U^\dagger(\bm k',\omega')\right\rangle =2\pi N_{\rm c} \delta_{\bm k,\bm k'}\delta(\omega-\omega')\bm S(\bm k,\omega),\\
\bm S(\bm k,\omega)=\bm G(\bm k,\omega)\bm Q(\omega)\bm G^\dagger(\bm k,\omega).
\end{gathered}
\label{eq:spectrum}
\end{equation}
It is a Hermitian, positive-semidefinite $4\times4$ matrix. Its four nonnegative eigenvalues quantify the fluctuation intensity of the corresponding eigenvectors. 

\subsection{Eigenvalues, gaps, and topology}
At fixed $(\bm k,\omega)$, we diagonalize the correlation matrix,
\begin{equation}
\bm S\bm v_n=s_n\bm v_n, \qquad s_1\leq s_2\leq s_3\leq s_4.
\end{equation}
Each orthonormal eigenvector $\bm v_n=(\mathv_{n,Ax},\mathv_{n,Ay},\mathv_{n,Bx},\mathv_{n,By})^{\T}$ defines a fluctuation pattern. Expanding the displacement as $\bm U=\sum_n a_n\bm v_n$, with $a_n=\bm v_n^\dagger\bm U$, yields
\begin{equation}
\left\langle a_m(\bm k,\omega)a_n^*(\bm k',\omega')\right\rangle =2\pi N_{\rm c} \delta_{\bm k,\bm k'}\delta(\omega-\omega')s_n(\bm k,\omega)\delta_{mn}.
\end{equation}
The eigenvectors therefore define mutually uncorrelated spectral components, with $s_n$ the corresponding fluctuation intensity. Unlike conventional lattice-band eigenvalues, the $s_n$ are not frequencies, energies, or relaxation rates.

\resultbox{The eigenvectors of $\bm S(\bm k, \omega)$ are modes of displacement fluctuations, not mechanical normal modes. Their eigenvalues measure fluctuation intensity rather than frequency or stiffness of mechanical modes.}

\begin{figure*}[t]
    \centering
    \includegraphics[width=0.99\linewidth]{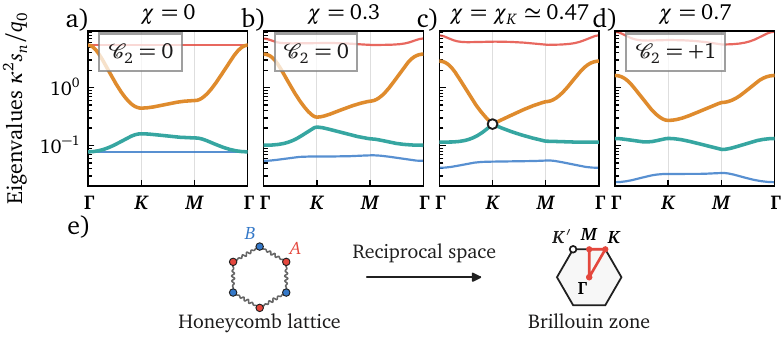}
    \caption{\textbf{Polarization-driven transition of the correlation bands.} \textbf{a--d)} Dimensionless eigenvalues $\kappa^2s_n/q_0$ along $\bm{\Gamma}\to \bm{K}\to \bm{M}\to\bm{\Gamma}$ for $\chi=0$, $0.30$, $\chi_K$, and $0.70$, respectively. Here $\chi_K\simeq0.47$ is the spectral polarization at which $s_2$ and $s_3$ touch at $\bm{K}$. \textbf{e)} Brillouin zone corresponding to the real-space lattice. In all panels, $\kappa/\delta=\kappa/\mu=2$, $\kappa\tau/\gamma=10$, and $\gamma\omega/\kappa=0.1$.}
    \label{fig:bands}
\end{figure*}

At fixed observation frequency $\omega$, the four eigenvalues $s_n(\bm k,\omega)$ can be plotted either as surfaces over $(k_x,k_y)$ or, more conveniently, along a path in the Brillouin zone, such as
\begin{equation}
\bm{\Gamma}\longrightarrow \bm{K}\longrightarrow \bm{M}\longrightarrow\bm{\Gamma},
\end{equation}
as shown in Fig.~\ref{fig:bands}(e). Panels (a)--(d) show the four eigenvalues along this path for several spectral polarizations at a fixed $\omega$. At the threshold $\chi=\chi_K\simeq 0.47$, the eigenvalues $s_2$ and $s_3$ touch at $\bm k=\bm{K}$ and separate again for $\chi>\chi_K$. We define the direct gap by
\begin{equation}
\Delta_m^{\rm dir}(\omega)=\min_{\bm k}\left[s_{m+1}(\bm k,\omega)-s_m(\bm k,\omega)\right].
\end{equation}
A positive direct gap isolates the rank-$m$ bulk subspace at every wavevector. At $\chi=\chi_K$, we find $\Delta_2^{\rm dir}=0$, while $\Delta_1^{\rm dir},\Delta_3^{\rm dir}>0$ at the frequency considered. This eigenvalue crossing is reminiscent of typical band topology. However, it does not establish nontrivial topology, so a topological invariant is required.

Whenever $\Delta_m^{\rm dir}>0$, the rank-$m$ subspace is isolated and carries the $m$th cumulative Chern number~\cite{PhysRevLett.51.51}
\begin{equation}
\begin{gathered}
\mathscr{C}_m(\omega)=\frac{1}{2\pi}\int_{\mathrm{BZ}}\mathcal F_m(\bm k,\omega) d^2k, \\
\mathcal F_m=\mathrm i\sum_{n=1}^{m}\left[\left(\partial_{k_x}\bm v_n\right)^\dagger\partial_{k_y}\bm v_n-\left(\partial_{k_y}\bm v_n\right)^\dagger\partial_{k_x}\bm v_n\right],
\end{gathered}
\end{equation}
where $\mathcal F_m$ is the Berry curvature of the correlation subspace below the $m$th direct gap. A nonzero $\mathscr{C}_m$ signals an obstruction to choosing a globally smooth and periodic basis of fluctuation patterns across the Brillouin zone. Because of its topological nature, $\mathscr{C}_m$ can change only when the direct gap closes, $\Delta_m^{\rm dir}=0$. This construction parallels the occupied-band Chern number of quantum band theory, but its physical interpretation is different: there is no Fermi level, particle occupation, or ground state.

Direct calculation from the eigenvectors corresponding to Fig.~\ref{fig:bands}(a)--(d) gives $\mathscr{C}_2=0$ for $0\leq\chi<\chi_K$, leaves it undefined at $\chi=\chi_K$, and gives $\mathscr{C}_2=+1$ in the final panel. The closing and reopening of the middle correlation gap therefore marks a genuine topological transition rather than an accidental band crossing. We now examine its origin and consequences.

% -------------------------------------------------------------------------
\section{Origin and phase diagram of the correlation topology}
\label{sec:topology}

\subsection{What is Chern-trivial and what is not}
\label{sec:trivial-ingredients}

The honeycomb stiffness matrix has $2\times2$  blocks,
\begin{equation}
\stiffness(\bm k)=\begin{pmatrix} \lambda_A\bm I_2&-\kappa\bm T(\bm k)\\
-\kappa\bm T^\dagger(\bm k)&\lambda_B\bm I_2 \end{pmatrix},\quad \lambda_{A,B}=\frac{3\kappa}{2}+\mu_{A,B}.
\label{eq:bl_stiffness}
\end{equation}
Using reciprocal coordinates $\bm k=u\bm b_1+\mathv\bm b_2$,
\begin{equation}
\bm T(u,\mathv)=\sum_i\bm P_i e^{\mathrm i\bm k\cdot\bm d_i}=\bm P_1+\bm P_2e^{\mathrm i2\pi u}+\bm P_3e^{\mathrm i2\pi \mathv},
\label{eq:bond_matrix}
\end{equation}
where $\bm P_j=\hat{\bm n}_j\hat{\bm n}_j^{\T}$ projects onto bond direction $j$, with
\begin{equation}
\hat{\bm n}_1=\begin{pmatrix}0\\
-1\end{pmatrix}, \quad \hat{\bm n}_2=\begin{pmatrix}\sqrt3/2\\
1/2\end{pmatrix}, \quad \hat{\bm n}_3=\begin{pmatrix}-\sqrt3/2\\
1/2\end{pmatrix},
\end{equation}
and $\bm d_1=\bm0$, $\bm d_2=\bm a_1$, and $\bm d_3=\bm a_2$ are the associated Bravais-cell shifts. Real reciprocal couplings imply
\begin{equation}
\stiffness(-\bm k)=\stiffness^*(\bm k).
\end{equation}
The Berry curvature associated with an isolated eigenspace of $\stiffness$ is therefore odd under $\bm k\to-\bm k$, so its Brillouin-zone integral vanishes. The same elastic matrix determines the propagator $\bm G$, which is independent of the noise and is also Chern-trivial in this case. The force spectrum $\bm Q$ is Chern-trivial because it is independent of $\bm k$. Hence, although we found that the correlation matrix $\bm S=\bm G\bm Q\bm G^\dagger$ of Eq.~\eqref{eq:spectrum} is topologically nontrivial, both $\bm G$ and $\bm Q$ are trivial.

Without chirality, $\chi=0$. Since Eq.~\eqref{eq:full_force_covariance} gives $\bm Q=q_0\bm I_4$, the correlation matrix is
\begin{equation}
\begin{aligned}
\bm S(\bm k,\omega)=q_0\bm G\bm G^\dagger=q_0\left[\stiffness^2(\bm k)+(\gamma\omega)^2\bm I_4\right]^{-1}.
\end{aligned}
\label{eq:nonchiral_S}
\end{equation}
Since $q_0(\omega)$ uniformly rescales all eigenvalues of $\bm S$, independently of $\bm k$, it cannot change its eigenspace topology. Hence, for $\chi=0$, the correlation matrix has vanishing Chern number. Nonzero spectral polarization is therefore required for the correlation eigenvectors to depart from those of the underlying reciprocal mechanical problem.

\resultbox{Chiral forcing makes $\bm S=\bm G\bm Q\bm G^\dagger$ topologically nontrivial although $\bm G$ and $\bm Q$ are individually Chern-trivial.}

We now investigate in more detail how exactly topology arises in our system by looking at the second gap.

\subsection{Closing of the gap at \texorpdfstring{$\bm{K}$}{K} and \texorpdfstring{$\bm{K}'$}{K'}}

We focus on the transition in Fig.~\ref{fig:bands}, where $s_2$ and $s_3$ cross and close the middle gap. We examine the other two gaps numerically in Sec.~\ref{sec:all-gap-numerics}.

Before addressing the topology of the transition, we first examine the gap closing at $\bm{K}$. We label displacement patterns by their sublattice and circular polarization:
\begin{equation}
\begin{gathered}
|A,+\rangle = \begin{pmatrix} \bm e_+\\
\bm 0_2 \end{pmatrix}, \qquad |A,-\rangle= \begin{pmatrix} \bm e_-\\
\bm 0_2 \end{pmatrix}, \\
|B,+\rangle = \begin{pmatrix} \bm 0_2\\
\bm e_+ \end{pmatrix}, \qquad |B,-\rangle= \begin{pmatrix} \bm 0_2\\
\bm e_- \end{pmatrix}.
\end{gathered}
\end{equation}

At $\bm{K}$, $|A,+\rangle$ and $|B,-\rangle$ are common eigenvectors of $\stiffness$ and $\bm Q$ because $\bm Q$ is diagonal in the circular basis, while the stiffness matrix satisfies
\begin{equation}
\bm T(\bm{K})\bm e_-=0, \qquad \bm T^\dagger(\bm{K})\bm e_+=0.
\end{equation}
The corresponding eigenvalues of $\bm S(\bm k=\bm{K},\omega)$ are (see Appendix~\ref{app:effective-correlation-theory}):
\begin{equation}
\begin{gathered}
\widetilde s_{A,+}(\bm{K})=q_0\frac{1+\chi}{D_A},\qquad \widetilde s_{B,-}(\bm{K})=q_0\frac{1-\chi}{D_B},\\
D_\alpha=\lambda_\alpha^2+(\gamma\omega)^2.
\end{gathered}
\end{equation}
The remaining two eigenvalues, corresponding to $s_1$ and $s_4$, do not enter this gap-closing argument.

Consider first $\delta>0$, for which $\mu_A>\mu_B$ and sublattice $A$ is more strongly pinned than sublattice $B$. Hence, $D_A>D_B$, and at $\chi=0$ one has $\widetilde s_{A,+}(\bm{K})<\widetilde s_{B,-}(\bm{K})$. Increasing $\chi>0$ transfers force power from the ``$-$'' sector to the ``$+$'' sector, causing $\widetilde s_{A,+}(\bm{K})$ to increase and $\widetilde s_{B,-}(\bm{K})$ to decrease. The two eigenvalues become equal at
\begin{equation}
\chi_K(\omega)=\frac{D_A-D_B}{D_A+D_B}=\frac{\lambda_A^2-\lambda_B^2}{\lambda_A^2+\lambda_B^2+2(\gamma\omega)^2}.
\end{equation}
This is the gap closing shown in Fig.~\ref{fig:bands}(c).

For the opposite spectral polarization, $\chi<0$, the crossing occurs at $\bm{K}'$ rather than at $\bm{K}$. Because
\begin{equation}
\bm T(\bm{K}')\bm e_+=0,\qquad \bm T^\dagger(\bm{K}')\bm e_-=0,
\end{equation}
the corresponding eigenvalues are
\begin{equation}
\widetilde s_{A,-}(\bm{K}')=q_0\frac{1-\chi}{D_A},\qquad \widetilde s_{B,+}(\bm{K}')=q_0\frac{1+\chi}{D_B}.
\end{equation}
For $\chi>0$, these levels move apart and the gap at $\bm{K}'$ increases. For $\chi<0$, they instead approach each other and cross at
\begin{equation}
\chi_{K'}=-\chi_K.
\end{equation}

If sublattice $B$ is instead more strongly pinned than $A$ ($\delta<0$), the roles of the two valleys are exchanged with the gap closing at $\bm{K}'$ for $\chi>0$ and at $\bm{K}$ for $\chi<0$.

\subsection{Topology at \texorpdfstring{$\bm{K}$}{K} and \texorpdfstring{$\bm{K}'$}{K'}}

As noted above, a gap closing is not topological merely because two eigenvalues become equal. To determine its topological effect, one must examine the eigenvectors around the touching. Appendix~\ref{app:effective-correlation-theory} provides the detailed derivation of the results that we now present.

Near either valley, $\bm{K}$ or $\bm{K}'$, the two central bands corresponding to $s_2$ and $s_3$ of $\bm S$ are found to be described by an effective $2\times2$ Hermitian matrix resembling a Dirac Hamiltonian:
\begin{equation}
\begin{aligned}
\bm S_{\mathrm{eff}}^{(\nu)}(\bm p) \simeq s_0^{(\nu)}\bm I_2+ \mathv_K\left(p_x\bm\sigma_x + \eta_\nu p_y\bm\sigma_y\right) + m_\nu\bm\sigma_z,
\end{aligned}
\end{equation}
where $\nu\in\{K,K'\}$, $\bm p=\bm k-\bm k_\nu$, $\bm k_K=\bm{K}$, $\bm k_{K'}=\bm{K}'$, $s_0^{(\nu)}=(s_2(\bm k_\nu)+s_3(\bm k_\nu))/2$ is the mean of the two levels, $\mathv_K>0$ is a finite coefficient, $\eta_K=-\eta_{K'}=1$, and $\bm\sigma_i$ are Pauli matrices:
\begin{equation}
\bm\sigma_x= \begin{pmatrix}0&1\\
1&0\end{pmatrix}, \quad \bm\sigma_y= \begin{pmatrix}0&-\mathrm i\\
\mathrm i&0\end{pmatrix}, \quad \bm\sigma_z= \begin{pmatrix}1&0\\
0&-1\end{pmatrix},
\end{equation}
and the Dirac ``mass'' $m_\nu$ controls the gap at valley $\nu$:
\begin{equation}
\begin{gathered}
m_K=m_0+m_\chi,\quad m_{K'}=m_0-m_\chi,\\
m_0=\frac{q_0}{2}\left(\frac{1}{D_A}-\frac{1}{D_B}\right),\quad m_\chi=\frac{q_0\chi}{2}\left(\frac{1}{D_A}+\frac{1}{D_B}\right).
\end{gathered}
\label{eq:valley_masses}
\end{equation}
The masses $m_K$ and $m_{K'}$ vanish at $\chi_K$ and $\chi_{K'}$, respectively, which closes the corresponding valley gap. A change of sign of one of these masses describes a local Dirac band inversion. Provided that no other gap closing occurs, such an inversion changes the Chern number by
\begin{equation}
\delta \mathscr{C}_2^{(\nu)}=+\frac12\operatorname{sgn}(J_\nu)\left[\operatorname{sgn}(m_\nu^{\rm after})-\operatorname{sgn}(m_\nu^{\rm before})\right],
\end{equation}
where $\delta\mathscr C_m=\mathscr C_m^{\rm after}-\mathscr C_m^{\rm before}$ denotes a parameter-induced change and $J_\nu$ is the local winding orientation. Since $J_K>0$ and $J_{K'}<0$, the mass inversion at $\bm{K}$ for increasing positive spectral polarization gives $\delta\mathscr{C}_2=+1$, while the corresponding inversion at $\bm{K}'$ for negative spectral polarization gives $\delta\mathscr{C}_2=-1$. Since the unpolarized system has $\mathscr{C}_2=0$, the reopened middle gap therefore has $\mathscr{C}_2=+1$ for $\chi>0$ and $\mathscr{C}_2=-1$ for $\chi<0$ immediately beyond the valley threshold $|\chi|>|\chi_K|$.

Equivalently, in the regime where no other gap closing occurs, these results can be summarized by
\begin{equation}
\mathscr{C}_2=+\frac12\left[\operatorname{sgn}(m_K)-\operatorname{sgn}(m_{K'})\right].
\end{equation}

The decomposition in Eq.~\eqref{eq:valley_masses} resembles the one found in Haldane's valley theory~\cite{haldane1988model}. In that model, a sublattice energy imbalance produces a common mass, while chiral hopping shifts the two valley masses in opposite directions. Here, sublattice pinning and chiral forcing play these respective roles through $m_0$ and $m_\chi$. For $|m_\chi|<|m_0|$, the masses have the same sign and the cones' opposite winding makes their Berry-curvature contributions cancel. For $|m_\chi|>|m_0|$, the masses have different sign and $\mathscr{C}_2=\operatorname{sgn}(\chi)$.

\subsection{Closing and topology of the gap at \texorpdfstring{$\bm{\Gamma}$}{Gamma}}

The previous valley analysis focused on $\bm k \simeq \bm{K}$ or $\bm k\simeq \bm{K}'$. As we will later see, another gap closing can occur at the center of the Brillouin zone $\bm{\Gamma}$. At this point,
\begin{equation}
\bm T(\bm{\Gamma})=\frac32\bm I_2,
\end{equation}
and the four eigenvalues are
\begin{equation}
\widetilde s_{j,\pm}(\bm{\Gamma})=\frac{q_\pm}{L_j}, \qquad j\in\{1,2\},
\end{equation}
with
\begin{equation}
\begin{gathered}
L_j = \Lambda_j^2+(\gamma\omega)^2,\qquad \Lambda_1<\Lambda_2,\\
\Lambda_{1,2} = \frac{\lambda_A+\lambda_B}{2} \mp \dfrac{1}{2}\sqrt{ \left( {\lambda_A-\lambda_B} \right)^2 + {9\kappa^2} },
\end{gathered}
\end{equation}
For positive spectral polarization, the middle levels are
\begin{equation}
\widetilde s_{2,+}=q_0\frac{1+\chi}{L_2}, \qquad \widetilde s_{1,-}=q_0\frac{1-\chi}{L_1}.
\end{equation}
They become degenerate at
\begin{equation}
\chi_\Gamma(\omega)=\frac{L_2-L_1}{L_2+L_1}=\frac{\Lambda_2^2-\Lambda_1^2}{\Lambda_2^2+\Lambda_1^2+2(\gamma\omega)^2}.
\end{equation}

The middle gap closes at $\chi=\pm\chi_\Gamma$, with $\chi_\Gamma>0$. We use the superscript $(\Gamma,+)$ to describe the transition when $\chi>0$ and $(\Gamma,-)$ when $\chi<0$. Near either transition, Appendix~\ref{app:effective-Gamma} gives
\begin{equation}
\begin{aligned}
\bm S_{\mathrm{eff}}^{(\Gamma,\pm)}(\bm p)\simeq{}&s_0^{(\Gamma,\pm)}\bm I_2 +\mathv^{(\Gamma,\pm)}(p_x\bm\sigma_x\mp p_y\bm\sigma_y)\\
&+m^{(\Gamma,\pm)}\bm\sigma_z,
\end{aligned}
\label{eq:Seff_Gamma}
\end{equation}
where $\bm p=\bm k-\bm{\Gamma}$, $s_0^{(\Gamma,\pm)}=(\widetilde s_{2,\pm}+\widetilde s_{1,\mp})/2$ is the mean of the two crossing levels, $\mathv^{(\Gamma,\pm)}>0$ are finite coefficients, and
\begin{equation}
m^{(\Gamma,\pm)}=\frac{q_0}{2}\left(\frac{1}{L_1}+\frac{1}{L_2}\right)(\pm\chi-\chi_\Gamma).
\end{equation}
Each mass vanishes at its corresponding signed threshold. As $|\chi|$ increases through $\chi_\Gamma$, the mass changes from negative to positive. The windings are $J^{(\Gamma,\pm)}=\mp[\mathv^{(\Gamma,\pm)}]^2$, so the inversions give $\delta\mathscr C_2=\mp1$ and restore $\mathscr C_2=0$. Hence, in these cases, the transitions happen via simple Dirac mass inversions at $\bm{\Gamma}$ rather than via Haldane's mechanism.

Provided that the valley closing precedes the $\bm{\Gamma}$ closing and that no other direct gap closing occurs, the middle-gap topology is given by
\begin{equation}
\mathscr{C}_2(\omega) = \begin{cases} \operatorname{sgn}[\chi(\omega)], & \displaystyle|\chi_K|<|\chi(\omega)|<\chi_\Gamma(\omega), \\[7pt]
0, & \text{otherwise} \end{cases}
\end{equation}
As $|\chi|$ increases with all other parameters fixed, an inversion at $\bm{K}$ or $\bm{K}'$ creates a phase with $\mathscr{C}_2=\pm 1$, which is later destroyed at $\bm{\Gamma}$.

\subsection{Numerical evaluation}
\subsubsection{With asymmetric pinning}

\begin{figure*}[t]
    \centering
    \includegraphics[width=0.99\linewidth]{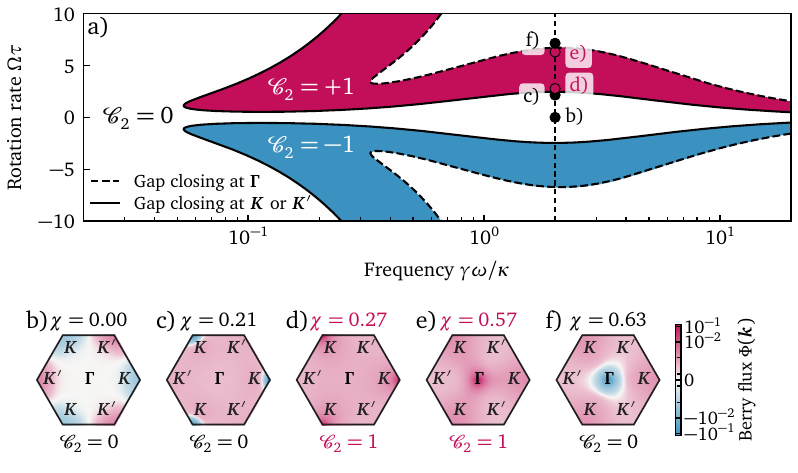}
    \caption{\textbf{Middle-gap topological diagram with asymmetric pinning.} \textbf{a)} Value of the middle-gap Chern number $\mathscr{C}_2$ as the frequency and dimensionless rotation rate are varied. The solid lines indicate the boundary $|\chi|=\chi_K$, while the dashed lines indicate $|\chi|=\chi_\Gamma$. \textbf{b--f)} Discretized Berry-curvature flux through one finite plaquette $\Phi(\bm k)$ at $\gamma\omega/\kappa=2$ as the spectral polarization is increased. The corresponding state points are indicated in \textbf{a)}. In all panels, $\kappa/\delta=\kappa/\mu=2$ and $\kappa\tau/\gamma=10$.}
    \label{fig:curvature}
\end{figure*}

Our results for $\delta\neq0$ are summarized in Fig.~\ref{fig:curvature}(a), which shows the middle-gap Chern number $\mathscr{C}_2$ as a function of the observation frequency $\omega$ and the dimensionless rotation rate $\Omega\tau$, the microscopic control parameter rather than the spectral polarization $\chi$. At $\Omega=0$, one has $\chi=0$, so the force spectrum is unpolarized and the middle-gap subspace has $\mathscr{C}_2=0$. Increasing $|\Omega|$ can drive the middle gap from $\mathscr{C}_2=0$ to $\mathscr{C}_2\neq0$ and subsequently back to $\mathscr{C}_2=0$. We insist here that we are varying $\omega\tau$ (the physically relevant tuning parameter) rather than $\chi$, which makes $\mathscr C_2$ slightly less transparent.

The solid phase boundaries satisfy $|\chi|=\chi_K$ and correspond to a closing of the middle correlation gap at $\bm{K}$ or $\bm{K}'$. The dashed boundaries satisfy $|\chi|=\chi_\Gamma$ and correspond to a gap closing at $\bm{\Gamma}$. Because
\begin{equation}
\chi(\omega,\Omega)=\frac{2\omega\Omega\tau^2}{1+\tau^2(\omega^2+\Omega^2)}
\end{equation}
is nonmonotonic in $|\Omega|$ at fixed $\omega$, the same threshold can be crossed twice. At sufficiently low frequency (\textit{e.g.} $\gamma\omega/\kappa=0.1$), the maximum accessible spectral polarization exceeds $\chi_K$ but remains below $\chi_\Gamma$. Increasing $\Omega>0$ then gives
\begin{equation}
\mathscr{C}_2:\qquad 0\xrightarrow{\bm{K}}+1\xrightarrow{\bm{K}}0.
\end{equation}
The first valley closing occurs while $\chi$ is increasing and creates the phase with $\mathscr{C}_2\neq0$, whereas the second occurs while $\chi$ is decreasing and restores $\mathscr{C}_2=0$ without any closing at $\bm{\Gamma}$. Along the displayed part of the vertical path in Fig.~\ref{fig:curvature}(a), for which $\gamma\omega/\kappa=2$, both thresholds are reached before $\chi$ attains its maximum, and the sequence is
\begin{equation}
\mathscr{C}_2:\qquad 0\xrightarrow{\bm{K}}+1\xrightarrow{\bm{\Gamma}}0.
\end{equation}
% Queue this wide figure early to keep it near its discussion.
\begin{figure*}[!t]
    \centering
    \includegraphics[width=0.99\linewidth]{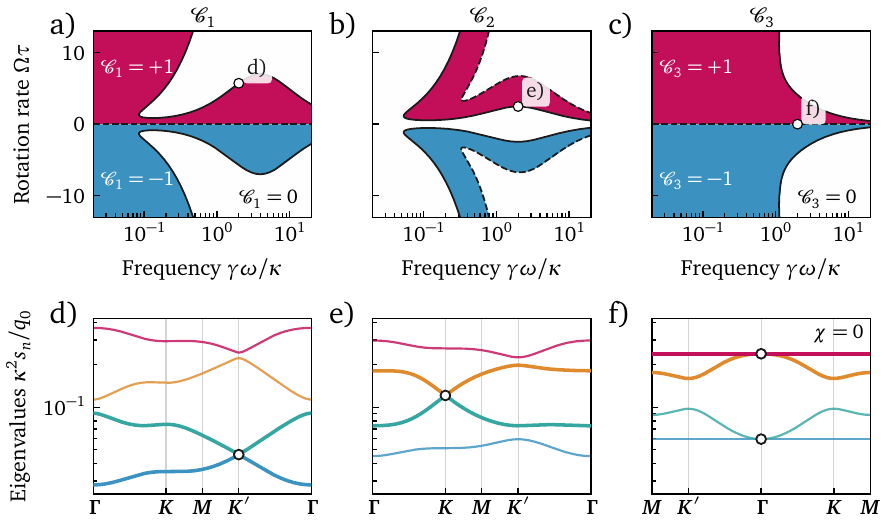}
    \caption{\textbf{Topology of all three correlation gaps.} \textbf{a--c)} Numerically evaluated gap Chern numbers $\mathscr{C}_1$, $\mathscr{C}_2$, and $\mathscr{C}_3$ in the observation-frequency and drive-parameter plane $(\gamma\omega/\kappa,\Omega\tau)$. Blue, white, and orange denote $-1$, $0$, and $+1$, respectively. Solid curves mark finite-$\chi$ gap closings, while dashed curves mark closings at $\bm{\Gamma}$; for the outer gaps in \textbf{a)} and \textbf{c)}, the latter occur along $\Omega=0$. \textbf{d--e)} Dimensionless correlation-band spectra $\kappa^2s_n/q_0$ along $\bm{\Gamma}\to \bm{K}\to \bm{M}\to \bm{K}'\to\bm{\Gamma}$ at the lower- and middle-gap valley transitions, respectively. \textbf{f)} Correlation bands at $\chi=0$ along $\bm{M}\to \bm{K}'\to\bm{\Gamma}\to \bm{K}\to \bm{M}$, with $\bm{\Gamma}$ centered and thicker lines for $s_3,s_4$. Both outer pairs touch quadratically at $\bm{\Gamma}$. Crossing from negative to positive $\chi$ reverses the chiral mass and changes $\mathscr C_1$ and $\mathscr C_3$ from $-1$ to $+1$. In \textbf{d--f)}, $\gamma\omega/\kappa=2$. In all panels, $\kappa/\delta=\kappa/\mu=2$ and $\kappa\tau/\gamma=10$.}
    \label{fig:all-gap-chern}
\end{figure*}

Continuing the same sweep to $\Omega\to\infty$, where $\chi$ decreases back to zero, crosses the two boundaries again and gives the full sequence
\begin{equation}
\mathscr{C}_2:\qquad 0\xrightarrow{\bm{K}}+1\xrightarrow{\bm{\Gamma}}0\xrightarrow{\bm{\Gamma}}+1\xrightarrow{\bm{K}}0.
\end{equation}

\begin{figure*}[!t]
    \centering
    \includegraphics[width=0.99\linewidth]{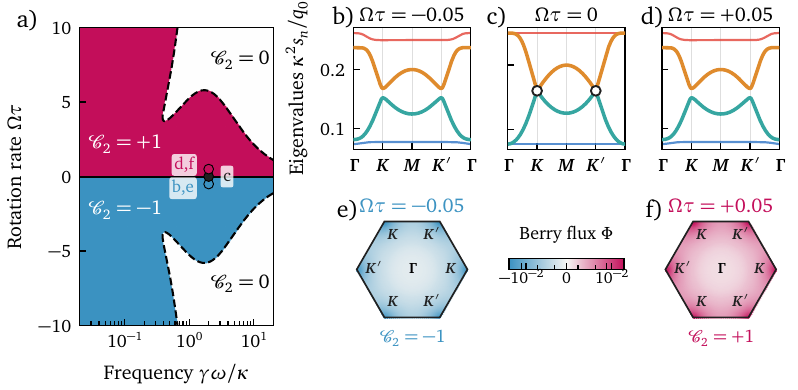}
    \caption{\textbf{Middle-gap topology without sublattice-asymmetric pinning.} \textbf{a)} Value of $\mathscr{C}_2$ as the frequency and dimensionless rotation rate are varied for $\mu_A=\mu_B=0$. The solid line indicates the simultaneous $\bm{K}/\bm{K}'$ middle-gap closing $|\chi|=\chi_K=0$, while the dashed lines indicate the $\bm{\Gamma}$-point boundary $|\chi|=\chi_\Gamma$. The selected state points at $\gamma\omega/\kappa=2$ correspond to the panels on the right. \textbf{b--d)} Dimensionless correlation eigenvalues $\kappa^2s_n/q_0$ along $\bm{\Gamma}\to \bm{K}\to \bm{M}\to \bm{K}'\to\bm{\Gamma}$. At $\Omega=0$, the middle correlation gap closes simultaneously at $\bm{K}$ and $\bm{K}'$. \textbf{e--f)} Middle-gap Berry flux for the two gapped states. In all panels, $\mu_A=\mu_B=0$ and $\kappa\tau/\gamma=10$.}
    \label{fig:unpinned}
\end{figure*}

Fig.~\ref{fig:curvature}(b)--(f) illustrates the first, increasing-$|\chi|$ half of this full sequence by showing the distribution of Berry flux over the Brillouin zone. We discretize the Brillouin zone on a mesh aligned with the primitive reciprocal vectors, containing $N_{\rm mesh}=N_1N_2$ wavevectors, with increments
\begin{equation}
\Delta\bm k_1=\frac{\bm b_1}{N_1},\qquad \Delta\bm k_2=\frac{\bm b_2}{N_2}.
\end{equation}
At each mesh point, the two eigenvectors below the middle gap are collected into
\begin{equation}
\bm V(\bm k)=\begin{pmatrix}\bm v_1(\bm k)&\bm v_2(\bm k)\end{pmatrix}.
\end{equation}
Following Refs.~\cite{fukui2005chern,hatsugai1993chern}, we define the Berry flux through an elementary plaquette starting at $\bm k$:
\begin{equation}
\begin{gathered}
\Phi(\bm k)\!=\!-\text{Arg}\!\left[U_1(\bm k)U_2(\bm k+\Delta\bm k_1) U_1^{-1}(\bm k+\Delta\bm k_2)U_2^{-1}(\bm k)\right]\!,\\
U_\nu(\bm k)=\frac{\det\left[\bm V^\dagger(\bm k)\bm V(\bm k+\Delta\bm k_\nu)\right]}{\left|\det\left[\bm V^\dagger(\bm k)\bm V(\bm k+\Delta\bm k_\nu)\right]\right|},\qquad \nu\in\{1,2\}.
\end{gathered}
\end{equation}
For a sufficiently fine mesh,
\begin{equation}
\Phi(\bm k)\simeq\int_{\rm plaquette}\mathcal F_2(\bm k')d^2k'\simeq\mathcal F_2(\bm k)\Delta A_{\bm k},
\end{equation}
where $\mathcal F_2(\bm k)$ is the Berry-curvature density and $\Delta A_{\bm k}$ is the plaquette area. The Chern number is obtained by summing these local fluxes over the full Brillouin zone,
\begin{equation}
\mathscr{C}_2=\frac{1}{2\pi}\sum_{\bm k\in\mathrm{BZ}}\Phi(\bm k).
\end{equation}
This sum is the discrete counterpart of $\mathscr{C}_2=(2\pi)^{-1}\int_{\mathrm{BZ}}\mathcal F_2(\bm k)d^2k$. The Berry-flux maps in Figs.~\ref{fig:curvature} and \ref{fig:unpinned} use $N_1=N_2=81$. Fig.~\ref{fig:curvature}(b)--(f) shows where the Berry curvature is concentrated. When $\mathscr{C}_2=0$, its positive and negative contributions cancel, whereas for $\mathscr{C}_2=+1$ the total flux is $+2\pi$. At $\chi=0$, the Berry curvature changes sign under wavevector reversal, so its magnitude is unchanged. Together with the lattice’s threefold rotational symmetry, this produces the sixfold pattern in $|\Phi|$. Finite chirality breaks the wavevector-reversal relation while preserving threefold rotations.

An important consequence of Fig.~\ref{fig:curvature} is that:
\resultbox{The Chern number depends on the frequency of observation $\omega$ of the fluctuations.}
We therefore use the term \emph{frequency-resolved topological phase}: the same steady state can exhibit different Chern numbers when observed at different frequencies. Equivalently, $\omega$ can be viewed as a synthetic coordinate, so that $\bm S(k_x,k_y,\omega)$ defines a three-dimensional parameter-space band structure~\cite{ozawa2019syntheticdimensions}. An isolated linear gap closing is then a synthetic Weyl point whose Berry charge determines the change in Chern number between adjacent $\omega$ slices~\cite{delplace2022berrychern}.

\subsubsection{Topology of all three correlation gaps}
\label{sec:all-gap-numerics}

The middle-gap invariant is only one part of the correlation-band topology. Fig.~\ref{fig:all-gap-chern} shows that the three gap invariants $\mathscr C_1$, $\mathscr C_2$ and $\mathscr C_3$ generally differ. In particular, the outer band pairs are degenerate at $\bm{\Gamma}$ when $\chi=0$, so $\mathscr{C}_1$ and $\mathscr{C}_3$ are undefined precisely on the zero-polarization line even though the middle gap remains open and $\mathscr{C}_2=0$. An infinitesimal positive spectral polarization gaps these quadratic touchings, as shown in Fig.~\ref{fig:all-gap-chern}(f) and derived in Appendix~\ref{app:quadratic-Gamma}. Unlike a Dirac point, where the bands split linearly away from the degeneracy, the splitting here is quadratic in momentum~\cite{chong2008effective}, and $|\delta\mathscr{C}_m| = 2$ upon crossing. Near $\bm{\Gamma}$, the corresponding reduced matrices are
\begin{equation}
\begin{aligned}
\bm S_{\mathrm{eff},j}^{(\Gamma)}(\bm p)\simeq{}&\left({q_0}/{L_j}+\beta_j\bm p^2\right)\bm I_2+m_j^{(\Gamma)}\bm\sigma_z\\
&+\beta_j\left[(p_x^2-p_y^2)\bm\sigma_x+2p_xp_y\bm\sigma_y\right],
\end{aligned}
\label{eq:quadratic_Gamma_main}
\end{equation}
where $\bm p=\bm k-\bm{\Gamma}$, $j=1$ labels the pair $(s_3,s_4)$ and $j=2$ the pair $(s_1,s_2)$. The real coefficients $\beta_j$ are given in Appendix~\ref{app:quadratic-Gamma}, and $m_j^{(\Gamma)}=q_0\chi/L_j$. The quadratic momentum terms wind twice around $\bm{\Gamma}$, while the mass reverses sign across $\chi=0$, accounting for the Chern-number jump of magnitude two. The resulting gaps carry
\begin{equation}
(\mathscr{C}_1,\mathscr{C}_2,\mathscr{C}_3)=(\pm 1,0,\pm 1), \qquad \chi = 0^\pm.
\end{equation}
Thus a phase that is ``trivial'' with respect to $\mathscr{C}_2$ can be topological in both outer gaps.

\subsubsection{Uniform pinning and the unpinned limit}

We next consider uniform pinning, $\mu_A=\mu_B=\mu$ ($\delta=0$), including the unpinned limit $\mu=0$. Although pinning no longer distinguishes the two sublattices, the honeycomb lattice remains non-Bravais, with two coupled sites per primitive unit cell. Accordingly, $\bm S$ remains a nontrivial $4\times4$ matrix. The nearest-neighbor central-force honeycomb also has an extensive floppy zero-stiffness band, but arbitrarily weak uniform pinning regularizes these modes.

Fig.~\ref{fig:unpinned}(a) displays two qualitative differences from asymmetric pinning. A middle-gap phase with $\mathscr{C}_2\neq0$ persists as $\Omega\to0^\pm$ and at arbitrarily small nonzero $\omega$. The $\mathscr{C}_2=0$ region is confined to intermediate spectral polarization and sufficiently large frequencies. This follows from
\begin{equation}
\chi_K(\omega)=\chi_{K'}(\omega)=0,\qquad \text{for}\qquad \mu_A=\mu_B,
\end{equation}
so crossing $\Omega=0$ simultaneously closes the middle gap at $\bm{K}$ and $\bm{K}'$, as shown in Fig.~\ref{fig:unpinned}(c), producing a jump $\delta\mathscr{C}_2=+2$. This corresponds to the limit of 0 sublattice imbalance in Haldane's model~\cite{haldane1988model}. Away from $\Omega=0$, successive middle-gap closings at $\bm{\Gamma}$ produce a $\mathscr{C}_2=0$ interval before the system reenters a phase with nonzero $\mathscr{C}_2$ at larger $|\Omega|$. Thus, for example, at $\gamma\omega/\kappa\simeq1$, sweeping $\Omega$ from $-\infty$ to $+\infty$ gives
\begin{equation}
\mathscr{C}_2:\quad \underbrace{-1\xrightarrow{\bm{\Gamma}}0\xrightarrow{\bm{\Gamma}}-1}_{\Omega:~~-\infty\to0^-}\xrightarrow{\bm{K},\bm{K}'}\underbrace{+1\xrightarrow{\bm{\Gamma}}0\xrightarrow{\bm{\Gamma}}+1}_{\Omega:~~0^+\to\infty}.
\end{equation}
At lower frequency, \textit{e.g.} $\gamma\omega/\kappa=0.1$, the sequence instead reduces to
\begin{equation}
\mathscr{C}_2:\qquad -1\xrightarrow{\bm{K},\bm{K}'}+1.
\end{equation}
This contrasts with asymmetric pinning, where $\mathscr{C}_2\xrightarrow{\omega\to 0}0$. Equal sublattice pinning therefore allows middle-gap topology at arbitrarily low nonzero frequencies. At exactly $\omega=0$, however, $\chi=0$ closes the middle gap, and the unpinned floppy band makes $\bm G$ singular.

Note that although the change $\delta \mathscr{C}_2$ at $\Omega=0$ is reminiscent of $\delta \mathscr{C}_{1,3}=2$ in Fig.~\ref{fig:all-gap-chern}, the underlying mechanisms are fundamentally different. In the unpinned case, the change by $2$ results from the vanishing asymmetry of the Haldane mechanism between $\bm{K}$ and $\bm{K}'$ for the middle bands. In the pinned case, it originates instead from a quadratic touching of the outer bands.

\section{Bulk-boundary correspondence}
\label{sec:bulk-boundary}

\subsection{Bulk--boundary correspondence for correlation bands}
\label{sec:boundary-meaning}

Bulk--boundary correspondence~\cite{asboth2016short} connects bulk band topology to modes localized at an edge or interface. For correlation bands, these modes are displacement-fluctuation patterns, and their eigenvalues $s$ measure fluctuation intensity.

At a physical edge, the lattice ends, and the exterior is vacuum, with no correlation bands and $\mathscr C_m^{\rm vac}=0$. If the bulk bands are separated by an interval containing no eigenvalues, a nonzero $\mathscr C_m$ requires boundary-localized bands connecting them across that gap. Each edge has at least $|\mathscr C_m|$ protected connections.

For an interface between two bulk regions, the gap must be shared by both spectra. A direct gap is enough to define each bulk Chern number, but eigenvalue ranges at different wavevectors can still overlap. At fixed $\omega$, define for each region $X\in\{\mathrm L,\mathrm R\}$
\begin{equation}
\begin{aligned}
I_m^X&=\left(\max_{\bm k}s_m^X(\bm k),\min_{\bm k}s_{m+1}^X(\bm k)\right),\\
\Delta_m^{\mathrm{ind},X}&=\min_{\bm k}s_{m+1}^X(\bm k)-\max_{\bm k}s_m^X(\bm k).
\end{aligned}
\end{equation}
When $\Delta_m^{\mathrm{ind},X}>0$, $I_m^X$ is an interval of width $\Delta_m^{\mathrm{ind},X}$ containing no bulk eigenvalue. For boundary localized modes, both intervals must overlap $I_m^{\mathrm L}\cap I_m^{\mathrm R}\neq\varnothing$ and have a different Chern number:
\begin{equation}
\Delta\mathscr C_m=\mathscr C_m^{\mathrm R}-\mathscr C_m^{\mathrm L}\neq 0.
\end{equation}
The magnitude of $\Delta \mathscr C_m$ gives the minimum number of protected boundary modes. Additional localized modes can exist without being topologically protected, although they will not cross the common gap.

Hence, by playing with the Chern numbers, it is possible to create localized interface modes of fluctuations. Since these modes are related to $\bm S$ and not to the deterministic propagator, they are not tied to transport.

\resultbox{A correlation boundary mode is a displacement-fluctuation pattern localized at an edge or interface. Its topology does not imply the existence of transport.}

\subsection{Bulk-boundary correspondence in a ribbon}

\begin{figure*}[t]
    \centering
    \includegraphics[width=0.99\linewidth]{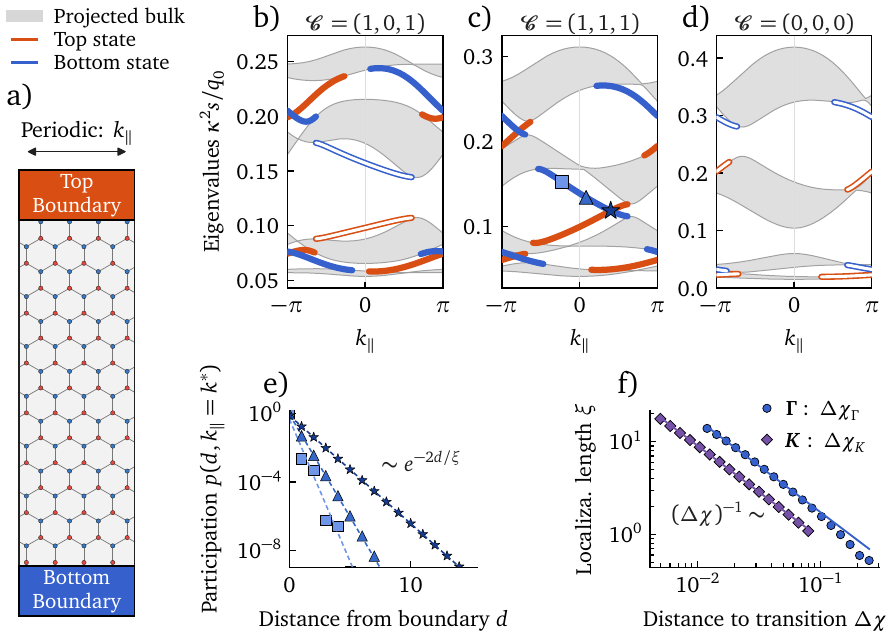}
    \caption{\textbf{Gap-resolved bulk--boundary correspondence at an edge-compensated termination.} \textbf{a)} Honeycomb ribbon. At each transverse edge, crossing-bond couplings are removed while their bulk diagonal stiffness contributions are retained, suppressing termination-induced floppy modes. \textbf{b--d)} Dimensionless ribbon eigenvalues $\kappa^2s/q_0$ for spectral polarizations $\chi=0.10$, $0.30$, and $0.75$, respectively. The corresponding Chern numbers are written above the panels. Gray regions are the projected bands of the infinite periodic bulk, obtained by varying $k_\perp$ at fixed $k_\parallel$. Solid orange and blue lines are protected top- and bottom-edge branches in gaps with $\mathscr{C}_m\neq0$; white lines with matching colored outlines are unprotected edge segments in gaps with $\mathscr{C}_m=0$. Lines stop when an edge state enters a projected bulk band. \textbf{e)} Row-resolved participation $p(d)$ of the three eigenvectors marked in \textbf{c)}. \textbf{f)} Localization length on approaching the $\bm{K}$ and $\bm{\Gamma}$ transitions, with $\Delta\chi_K=|\chi-\chi_K|$ and $\Delta\chi_\Gamma=|\chi-\chi_\Gamma|$. In all numerical panels, $\kappa/\delta=\kappa/\mu=2$, the ribbon contains $N_y=32$ transverse unit-cell rows, $\kappa\tau/\gamma=10$, and $\gamma\omega/\kappa=2$.}
    \label{fig:bulk-boundary}
\end{figure*}

We consider a ribbon geometry, translationally invariant along one direction (periodic in our case) but terminated transversely, as illustrated in Fig.~\ref{fig:bulk-boundary}(a). The displacement patterns therefore decompose into sectors labeled by $k_\parallel$. At each $k_\parallel$, we define
\begin{equation}
\begin{gathered}
\bm G_{\rm rib}(k_\parallel,\omega) = \left[\stiffness_{\rm rib}(k_\parallel)-\mathrm i\gamma\omega\bm I\right]^{-1}, \\
\bm S_{\rm rib} =\bm G_{\rm rib}\bm Q_{\rm rib}\bm G_{\rm rib}^{\dagger}.
\end{gathered}
\end{equation}
Unlike in the fully periodic bulk, translational invariance is lost in the transverse direction, so there is no transverse wavevector $k_\perp$ for the ribbon. The transverse unit-cell rows must therefore be retained explicitly in real space. Since each unit cell contains two sublattice sites, $A$ and $B$, with two displacement components per site, each transverse row contributes four degrees of freedom. Consequently, for a ribbon containing $N_y$ unit-cell rows in the transverse direction, $\stiffness_{\rm rib}$, $\bm G_{\rm rib}$, $\bm Q_{\rm rib}$, and $\bm S_{\rm rib}$ are $4N_y\times4N_y$ matrices at fixed $k_\parallel$, rather than the $4\times4$ matrices of the fully periodic bulk.

The matrix $\bm S_{\rm rib}$ can then be diagonalized independently at each $k_\parallel$ to obtain the eigenvalues $s_\ell(k_\parallel)$ and corresponding normalized eigenvectors $\bm v_\ell(k_\parallel)$, with $\ell=1,2,\dots,4N_y$. The observation frequency is held fixed and suppressed here when unambiguous. Each eigenvector has $4N_y$ components: for each of the $N_y$ transverse unit-cell rows, there are two sublattice components, $\beta\in\{A,B\}$, and two  components, $p\in\{x,y\}$. Projection over all $k_\perp$ produces dense bulk continua at each $k_\parallel$, separated by three gaps.

As $k_\parallel$ varies, edge-localized eigenvalues form branches that can connect the bulk bands on opposite sides of a gap. This connectivity is called boundary spectral flow. For a fixed interface orientation, branches crossing the gap in opposite directions count with opposite signs, and their net number is $\Delta\mathscr C_m$. For an interface between a topological bulk and vacuum, $|\Delta\mathscr C_m|=|\mathscr C_m|$. Moreover, because $s_\ell$ measures fluctuation intensity, $\partial s_\ell/\partial k_\parallel$ is not a group velocity, so spectral flow does not imply topologically protected transport~\footnote{This does not mean that the modes are not propagating.}.

The spatial distribution of an eigenvector $\ell$ across the ribbon can therefore be characterized by its row-resolved intensity
\begin{equation}
    \begin{gathered}
p_\ell(d,k_\parallel,\omega)=\sum_{\beta\in\{A,B\}}\sum_{p\in\{x,y\}}\left|\mathv_{\ell,d\beta p}(k_\parallel, \omega)\right|^2,\\
\sum_{d=1}^{N_y}p_\ell(d,k_\parallel,\omega)=1,
    \end{gathered}
\end{equation}
where $d$ labels the transverse unit-cell row.  $p_\ell(d,k_\parallel,\omega)$ measures the relative participation of eigenvector $\ell$ in transverse row $d$, at a given $k_\parallel$ and $\omega$.

Fig.~\ref{fig:bulk-boundary}(b)--(d) shows the eigenvalues of $\bm S_{\rm rib}$ as functions of $k_\parallel$ for several parameter values. The gray regions represent the bulk eigenvalues of the $4\times4$ problem, projected from $s(\bm k)$ to $s(k_\parallel)$, while the boundary-localized modes are computed from the ribbon eigenvalue problem at finite $N_y$. To identify these eigenvectors, we compute their total intensity within the three unit-cell rows closest to each transverse boundary. Denoting these weights by
\begin{equation}
\begin{gathered}
w_{\rm top}^{(\ell)}(k_\parallel)=\sum_{d\in\text{top three rows}}p_\ell(d,k_\parallel), \\
w_{\rm bottom}^{(\ell)}(k_\parallel)=\sum_{d\in\text{bottom three rows}}p_\ell(d,k_\parallel),
\end{gathered}
\end{equation}
we classify an eigenvector $\ell$ as strongly boundary-localized when $w_{\rm top}^{(\ell)}+w_{\rm bottom}^{(\ell)}>0.5$ (which is arbitrary). Eigenvectors with little boundary weight are supported primarily in the ribbon interior and are thus bulk-like. Edge-localized eigenvalues outside the projected bulk spectrum are shown in the color of the edge on which their eigenvectors are localized. Filled curves denote edge branches spanning an entire bulk gap and are therefore topologically protected. Unfilled curves enter a gap and reconnect to the same bulk band. Although edge-localized, these states are not topologically protected.

For example, in Fig.~\ref{fig:bulk-boundary}(b), the lower and upper gaps have $\mathscr{C}_1=\mathscr{C}_3=1$, while $\mathscr{C}_2=0$. The vacuum outside either edge is Chern-trivial. The mismatch therefore requires one protected traversal across the lower and upper gaps at each boundary but none across the middle gap, whose edge states return to the same bulk band. In Fig.~\ref{fig:bulk-boundary}(c), all three Chern numbers are nonzero, $|\mathscr{C}_m|=1$, and each gap contains two traversing branches, one at each boundary. In Fig.~\ref{fig:bulk-boundary}(d), $|\mathscr{C}_m|=0$ for all three gaps, and every edge branch reconnects to the same bulk continuum.

\begin{figure*}[t]
    \centering
    \includegraphics[width=0.99\linewidth]{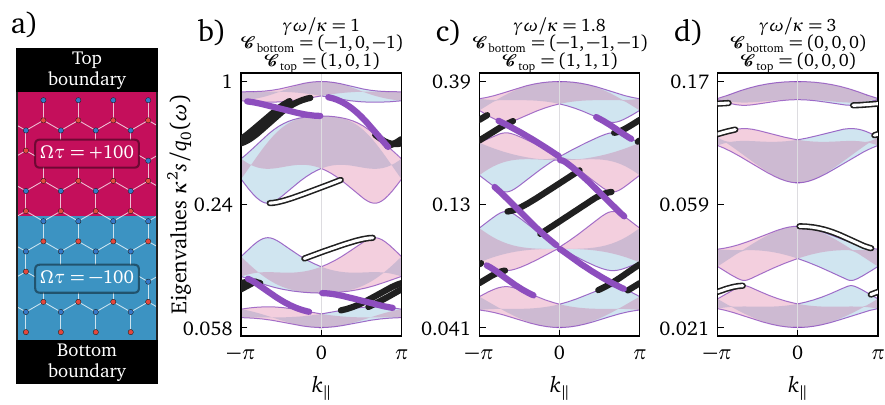}
    \caption{\textbf{Frequency-dependent bulk--boundary correspondence at an activity interface.} \textbf{a)} Mechanically uniform ribbon with $\Omega\tau=-100$ and $+100$ in its bottom and top halves, respectively. \textbf{b--d)} Correlation spectra at $\gamma\omega/\kappa=1.0$, $1.8$, and $3.0$. The red and blue regions are the projected bulk bands. Solid purple and black lines denote topological branches localized at the activity interface and mechanical boundaries, respectively; white lines with black outlines denote localized but topologically unprotected branches outside the projected bulk bands, following the convention of Fig.~\ref{fig:bulk-boundary}. The microscopic parameters $D_a$, $\Omega$, and $\tau$ are fixed in all panels; eigenvalues are shown as $\kappa^2s/q_0(\omega)$, so the panel-dependent overall force-spectrum scale is divided out. The two outer mechanical edges use the same edge-compensated termination as Fig.~\ref{fig:bulk-boundary}. Here $N_y=36$, $\kappa/\delta=\kappa/\mu=2$, and $\kappa\tau/\gamma=10$.}
    \label{fig:chern-jump-two}
\end{figure*}

As expected from bulk--boundary correspondence, the edge states are exponentially localized, with $p(d)\sim e^{-2d/\xi}$, where $\xi$ is the localization length measured in unit-cell rows. This behavior is shown in Fig.~\ref{fig:bulk-boundary}(e), which displays the participation profiles of three eigenvectors associated with the eigenvalues marked in Fig.~\ref{fig:bulk-boundary}(c). Near both transitions, Eqs.~\eqref{eq:app_Seff_K_final} and \eqref{eq:app_Seff_Gamma} reduce the correlation spectrum to a massive Dirac matrix whose mass vanishes linearly, $|m|\propto\Delta\chi$, with $\Delta\chi=|\chi-\chi_c|$ and $\chi_c=\chi_K$ or $\chi_\Gamma$. Because the decay length of a Dirac boundary mode is inversely proportional to its ``mass''~\cite{dasbiswas2018topological}, the localization length diverges as $\xi\sim|m|^{-1}\sim(\Delta\chi)^{-1}$. The edge states therefore delocalize progressively and merge with the bulk at the transition, as shown in Fig.~\ref{fig:bulk-boundary}(f).

\subsection{Activity wall in a mechanically uniform strip}

An interface need not be a region where the system terminates. For instance, imposing $\Omega\tau=-100$ on one half of the ribbon and $\Omega\tau=+100$ on the other creates an activity-induced interface, as shown in Fig.~\ref{fig:chern-jump-two}(a). Fig.~\ref{fig:chern-jump-two}(b)--(d) shows the corresponding spectra at different observation frequencies $\omega$.

We denote the three cumulative gap invariants by $\lpmb{\mathscr{C}}=(\mathscr{C}_1,\mathscr{C}_2,\mathscr{C}_3)$. The absolute value should be understood componentwise: $|\lpmb{\mathscr{C}}|=(|\mathscr{C}_1|,|\mathscr{C}_2|,|\mathscr{C}_3|)$. In Fig.~\ref{fig:chern-jump-two}(b),
\begin{equation}
|\lpmb{\mathscr{C}}_{\rm top}-\lpmb{\mathscr{C}}_{\rm bottom}|=(2,0,2),
\end{equation}
yielding four topological branches localized at the activity-induced interface. Four additional branches, shown in black, are localized at the mechanical boundaries because
\begin{equation}
\begin{gathered}
|\lpmb{\mathscr{C}}_{\rm top/bottom}-\lpmb{\mathscr{C}}_{\rm vacuum}|=(1,0,1).
\end{gathered}
\end{equation}
In Fig.~\ref{fig:chern-jump-two}(c), all three Chern mismatches are nonzero and
\begin{equation}
|\lpmb{\mathscr{C}}_{\rm top}-\lpmb{\mathscr{C}}_{\rm bottom}|=(2,2,2).
\end{equation}
Each gap therefore contains two purple interface branches and one black branch at each mechanical boundary. At the higher frequency shown in Fig.~\ref{fig:chern-jump-two}(d), all gap Chern numbers vanish, and the edge states are not topologically protected.

We insist that in Fig.~\ref{fig:chern-jump-two}(b-d), we only changed the frequency of observation $\omega$.

\subsection{Closing the activity interface into a loop}

\begin{figure*}[t]
    \centering
    \includegraphics[width=0.99\linewidth]{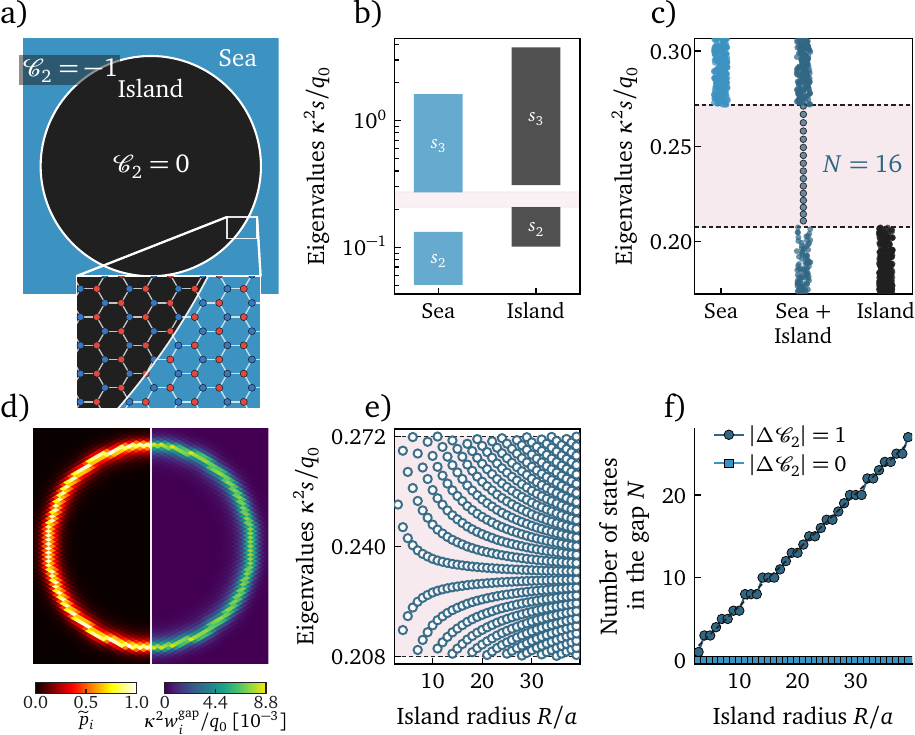}
    \caption{\textbf{Bulk--boundary correspondence at a closed activity interface.} \textbf{a)} A circular island of radius $R/a\simeq24$, with spectral polarization $\chi_{\rm island}=-0.30$ and $\mathscr{C}_2=0$, embedded in a sea with $\chi_{\rm sea}=-0.70$ and $\mathscr{C}_2=-1$; $a$ is the Bravais lattice constant defined in Sec.~\ref{sec:spectral-objects}. \textbf{b)} Ranges of the homogeneous dimensionless bands $\kappa^2s_2/q_0$ and $\kappa^2s_3/q_0$ in the sea and island. \textbf{c)} Dimensionless eigenvalues near that gap for a uniform sea, the sea--island system, and a uniform island. There are $N$ eigenvalues inside the common gap. \textbf{d)} Real-space weight of the in-gap modes: the left half shows their maximum-normalized participation $\widetilde p_i$, while the right half shows the dimensionless local spectral power $\kappa^2w_i^{\rm gap}/q_0$ carried by the same modes, on an independent color scale. \textbf{e)} Discrete in-gap eigenvalues as the island radius $R/a$ increases. \textbf{f)} Number $N$ of in-gap eigenvalues versus $R/a$ for the topological interface, $|\Delta\mathscr{C}_2|=1$, and for a control interface with $\chi_{\rm control}=-0.55$ on the island, for which $|\Delta\mathscr{C}_2|=0$. The calculation keeps $q_0$ spatially uniform; realizing the two stated values of $|\chi|$ with the microscopic Ornstein--Uhlenbeck process generally requires choosing the local $D_a$ together with $\Omega$. All panels use a periodic $92\times92$-cell system, $\kappa/\delta=\kappa/\mu=2$, and $\gamma\omega/\kappa=0.1$.}
    \label{fig:closed-island}
\end{figure*}

Finally, we break translational symmetry completely by considering a periodic system containing a circular island of different activity:
\begin{equation}
\chi(\bm r)= \begin{cases} \chi_{\rm island}=-0.30, & |\bm r|<R,\\
\chi_{\rm sea}=-0.70, & |\bm r|>R, \end{cases}
\end{equation}
where $R$ is the island radius, as illustrated in Fig.~\ref{fig:closed-island}(a). For this problem, no simplification can be done, and the full matrix $\bm S$ of size $4N_xN_y\times 4 N_x N_y$ has to be diagonalized. At the frequency considered, the island and surrounding sea have $\mathscr{C}_2=0$ and $\mathscr{C}_2=-1$, respectively. The two homogeneous systems share the open middle gap shown in Fig.~\ref{fig:closed-island}(b). Their circular interface therefore has a mismatch $|\Delta\mathscr{C}_2|=1$ and supports the in-gap boundary modes shown in Fig.~\ref{fig:closed-island}(c). Fig.~\ref{fig:closed-island}(d) confirms their localization. Its left half shows the participation of in-gap modes $\widetilde p_i$ at unit cell $i$, while its right half shows their local spectral power $w_i^{\rm gap}$:
\begin{equation}
\begin{gathered}
p_i=\sum_{\ell\in\mathrm{gap}} \sum_{\beta\in\{A,B\}}\sum_{p\in\{x,y\}} \left|\mathv_{\ell,i\beta p}\right|^2, \qquad \widetilde p_i=\frac{p_i}{\max_j p_j},\\
w_i^{\rm gap}=\sum_{\ell\in\mathrm{gap}}s_\ell\sum_{\beta\in\{A,B\}}\sum_{p\in\{x,y\}}\left|\mathv_{\ell,i\beta p}\right|^2.
\end{gathered}
\end{equation}
As shown in Fig.~\ref{fig:closed-island}(e)--(f), the number of in-gap states increases with the island radius.

\begin{figure*}[t]
    \centering
    \includegraphics[width=0.99\linewidth]{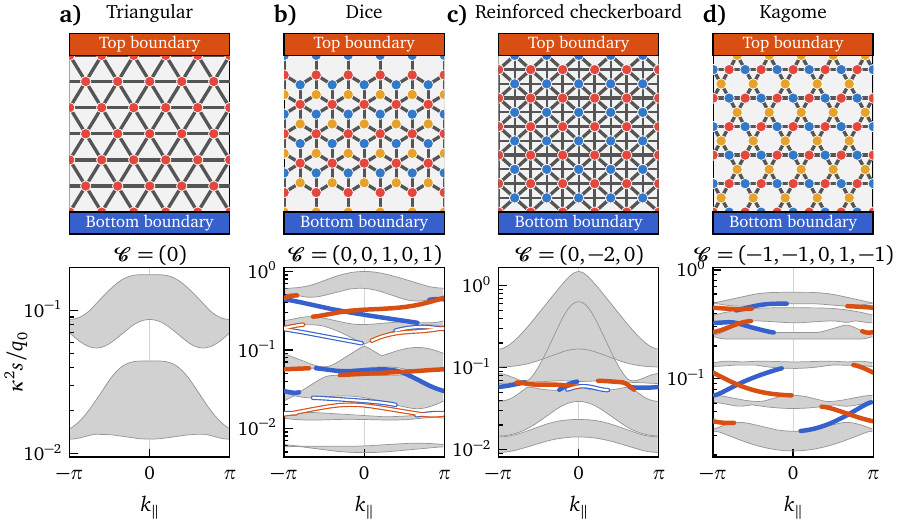}
    \caption{\textbf{Complete correlation spectra for four reciprocal lattices.} Dimensionless bands $\kappa^2s/q_0$ for lattices with edge-compensated top and bottom terminations: \textbf{a)} unpinned triangular lattice, \textbf{b)} pinned dice lattice, \textbf{c)} pinned reinforced checkerboard, and \textbf{d)} unpinned kagome lattice. Crossing-bond couplings are removed while the bulk diagonal stiffness blocks are retained. Gray projected bulk bands and colored or white-core boundary branches follow the conventions of Fig.~\ref{fig:bulk-boundary}. In every panel, $\chi$ denotes the spectral polarization. The triangular lattice has equal nearest-neighbor springs, $\mu=0$, $\gamma\omega/\kappa=3.00$, $\chi=0.60$, and $N_y=26$. The dice lattice has equal nearest-neighbor springs and $(\mu_0,\mu_1,\mu_2)/\kappa=(0,4,5/4)$; $\gamma\omega/\kappa=3/4$, $\chi=4/5$, and $N_y=24$. For the checkerboard, $(\mu_A,\mu_B)/\kappa=(3/2,1/2)$, all four diagonal $A$--$B$ and both axial $A$--$A$ spring constants are $\kappa$, while both axial $B$--$B$ spring constants are $\kappa/2$; $\gamma\omega/\kappa=1/4$, $\chi=2/5$, and $N_y=28$. The kagome lattice has equal nearest-neighbor springs, $\mu_\alpha=0$, $\gamma\omega/\kappa=1.50$, $\chi=0.40$, and $N_y=24$.}
    \label{fig:other-lattices}
\end{figure*}

\section{Discussion and outlook}
\label{sec:discussion}

We have shown that Chern topology can be carried by fluctuations rather than by the deterministic dynamics itself. In reciprocal, Chern-trivial elastic lattices driven by local chiral active noise, the frequency-resolved displacement spectrum acquires nonzero Chern bands. This mechanism produces topological transitions controlled by the drive statistics and the observation frequency and, through bulk--boundary correspondence, boundary-localized spectral flow in fluctuation intensity without implying unidirectional mechanical transport. Because the construction relies on the structure of the correlation operator rather than on a specific deterministic topological band model, it suggests a broader program in which correlation spectra are treated as topological objects in their own right, including other lattice geometries, stochastic drivings, interfaces, and higher-dimensional systems. We now discuss the ingredients essential to this phenomenon and its extension beyond the honeycomb lattice.

\subsection{Design principles beyond the honeycomb lattice}
\label{sec:discussion-design}

The honeycomb result illustrates how reciprocal elastic interactions and local chiral forcing can produce Chern-nontrivial fluctuations. Its topological onset is explained by the valley mass inversion above. For another lattice, topology must be assessed from the full correlation matrix $\bm S=\bm G\bm Q\bm G^\dagger$, whose eigenvectors depend on both the elastic geometry and the random noise polarization.

Nonzero spectral polarization and an isolated correlation subspace are necessary in the class of models considered here, but they do not guarantee a nonzero Chern number. The lattice structure can impose further constraints on the correlation eigenvectors, as illustrated by the Chern-trivial triangular lattice below.

\subsection{Other lattices and the role of pinning}
\label{sec:discussion-lattices}

Fig.~\ref{fig:other-lattices} applies the same chiral forcing to four reciprocal lattices with edge-compensated top and bottom terminations. Fig.~\ref{fig:other-lattices}(a) shows an unpinned triangular lattice with one site per unit cell and two correlation bands. Its gap is Chern-trivial, $\lpmb{\mathscr{C}}=(0)$, because the symmetries of this central-force model prevent a nonzero Chern number. Fig.~\ref{fig:other-lattices}(b)--(d) shows several non-honeycomb lattices with Chern-nontrivial fluctuations.

Pinning is not necessary for topology. Uniform pinning nevertheless regularizes long-wavelength motion, while sublattice-dependent pinning can separate bands and improve visualization, as in the dice and checkerboard examples. Unpinned lattices may contain zero-stiffness floppy modes, yielding to a nonstationary process driven by the random noise. In this case, most results must be understood as having vanishing pinning. The limit $\omega\to0$ requires separate care because acoustic or floppy modes may make the propagator diverge.

\subsection{Is chiral forcing necessary?}
\label{sec:discussion-chirality}

For the reciprocal propagator and local stochastic forcing considered here, chirality in the forcing is necessary. When $\chi=0$, $\bm Q=q_0\bm I$, and Eq.~\eqref{eq:nonchiral_S} makes $\bm S$ a scalar function of the reciprocal stiffness matrix $\stiffness$. The correlation matrix then satisfies $\bm S(-\bm k,\omega)=\bm S^*(\bm k,\omega)$, which makes the Berry curvature odd in momentum and forces the Chern number of every isolated subspace to vanish.

The particular chiral Ornstein--Uhlenbeck process used here is not necessary in a more general system. A nonreciprocal bath, sublattice-dependent circular forcing, or a complex momentum-dependent spectrum $\bm Q(\bm k,\omega)$ may also generate topology. For the last possibility, the bulk--boundary argument above additionally requires $\bm Q^{-1}$ to remain spatially local. Conversely, chirality may enter the propagator $\bm G$ through gyroscopic couplings, odd elasticity, or other nonreciprocal elements~\cite{nash2015gyroscopic,souslov2017topological,scheibner2020odd,fossati2024odd}. Achiral forcing can then reveal topology already encoded by the propagator, and $\bm S$ may display the same nontrivial topology as in the present model, although it would also exist in the deterministic dynamics.

\subsection{Correlation edge modes are not automatically propagating modes}
\label{sec:discussion-propagation}

For a ribbon geometry, the slope $\partial s/\partial k_\parallel$ is not a group velocity. Because $s$ quantifies fluctuations, the bulk--boundary correspondence does not itself guarantee that a wavepacket travels unidirectionally along the boundary.

Propagation of a coherent perturbation at frequency $\omega$ is instead controlled by the spatial Green function,
\begin{equation}
\begin{aligned}
\bm U_n(\omega)={}&\frac{1}{N_{\rm c}}\sum_{\bm k}e^{\mathrm i\bm k\cdot\bm R_n}\bm G(\bm k,\omega)\bm F_{\mathrm{ext}}(\bm k,\omega).
\end{aligned}
\end{equation}
The poles, damping, and mechanical dispersion of $\bm G$ determine the spreading and attenuation of perturbations. In the model studied here, $\bm G$ is reciprocal and mechanically Chern-trivial, so the Chern number of $\bm S$ cannot be interpreted as protected one-way transport. This contrasts with mechanical or active systems whose dynamical bands themselves carry topology and support propagating edge waves~\cite{susstrunk2015phononic,nash2015gyroscopic,souslov2017topological}.

This does not make the boundary correlations static, and a localized stochastic perturbation can still spread through the reciprocal propagator $\bm G$. The resulting propagation, however, is not robustly fixed by a topological invariant.

\subsection{Dimensionality}

In three dimensions, local rotation is specified by a pseudovector $\bm\Omega=\Omega\hat{\bm c}$ rather than by a pseudoscalar. The 3D extension of Eq.~\eqref{eq:ou} is
\begin{equation}
\tau\dot{\bm f}_{n\alpha}=-\bm f_{n\alpha}+\tau\bm\Omega\times\bm f_{n\alpha}+\sqrt{2D_a\tau} \bm\xi_{n\alpha}.
\end{equation}
The cross product simply rotates the force around $\hat{\bm c}$. Related three-dimensional chiral active models produce helical motion and anisotropic correlations~\cite{lettermann2025three,kuroda2025singular}. Choosing $\hat{\bm c}=\hat{\bm z}$, the force spectrum takes the explicit form
\begin{equation}
\begin{gathered}
\bm q(\omega)=\frac{1}{2}\begin{pmatrix}q_++q_-&-\mathrm i(q_+-q_-)&0\\
\mathrm i(q_+-q_-)&q_++q_-&0\\
0&0&2q_z\end{pmatrix},\\
q_z=\frac{2D_a\tau}{1+\tau^2\omega^2},
\end{gathered}
\end{equation}
where $q_\pm$ are given by Eq.~\eqref{eq:q_def}. Its three eigenvalues are $q_+$, $q_-$, and $q_z$ and are therefore nonnegative. The rotation axis leaves the longitudinal component unpolarized while splitting the two transverse circular polarizations.

Consider $\bm\Omega=\Omega\hat{\bm z}$ and a stack of two-dimensional lattices coupled along $z$. At each fixed wavevector $k_z$, the matrix $\bm S(k_x,k_y,k_z,\omega)$ defines a two-dimensional correlation-band problem with slice Chern number $\mathscr{C}(k_z)$. If the interlayer coupling is weak enough to preserve the correlation gap, every slice retains the nonzero Chern number of an isolated layer. The boundary bands of the layers then combine into fluctuation bands localized on the side surfaces of the stack. Increasing the interlayer coupling can make $\mathscr{C}(k_z)$ depend on $k_z$, but this integer can change only when a slice gap closes. An isolated closing in the full three-dimensional wavevector space is a Weyl point that carries quantized Berry flux and can produce surface correlation arcs. Mechanical Weyl points and surface arcs have already been realized in metamaterials~\cite{rocklin2016mechanical,shi2019elastic}.

\begin{acknowledgments}

I thank Yuta Kuroda for introducing me to the field of topological active matter. Part of this work was initiated during the E. Majorana-WE-Heraeus Workshop on Hyperuniformity in Erice, Italy.

This research received no external funding.

\end{acknowledgments}

\section*{Data availability}

The source code used to generate the figures is available at \href{https://github.com/Syrocco/topology}{github.com/Syrocco/topology}.

\onecolumngrid
\appendix

\section{Two-band theory of the middle correlation-gap closings}
\label{app:effective-correlation-theory}

At fixed observation frequency $\omega$, we derive the two-band theories for $s_2$ and $s_3$. We first establish a projection formula valid at any symmetry point where the stiffness $\stiffness$ and force spectrum $\bm Q$ share the two crossing eigenvectors. We then apply it at $\bm{K}$, $\bm{K}'$, and $\bm{\Gamma}$ and infer the changes of $\mathscr{C}_2$ from the masses and momentum-space windings.

For simplicity, we consider the regime studied in the main text: $\lambda_A>\lambda_B$, and, as $\chi>0$ increases from zero, the valley closing at $\bm{K}$ precedes the closing at $\bm{\Gamma}$. Negative $\chi$ exchanges $\bm{K}$ and $\bm{K}'$ and reverses the circular polarizations. For $\delta<0$, the roles of the two valleys are also exchanged. We suppress the fixed $\omega$ argument and define
\begin{equation}
c_\omega=\gamma\omega,\qquad t_\kappa=\frac{3\kappa}{2},\qquad D_\alpha=\lambda_\alpha^2+c_\omega^2,\qquad q_\pm=q_0(1\pm\chi).
\end{equation}
The nearest-neighbor distance is the unit of length.

\subsection{General two-band reduction near an isolated crossing}
\label{app:projection-general}

Let $\bm k_c$ be a crossing point and let $|u\rangle$ and $|v\rangle$ be two orthonormal common eigenvectors of $\stiffness_c\equiv\stiffness(\bm k_c)$ and $\bm Q$:
\begin{equation}
\stiffness_c|r\rangle=\lambda_r|r\rangle,\qquad \bm Q|r\rangle=q_r|r\rangle,\qquad r\in\{u,v\}.
\end{equation}
Writing $D_r=\lambda_r^2+c_\omega^2$, Eq.~\eqref{eq:spectrum} gives $\bm S(\bm k_c)|r\rangle=(q_r/D_r)|r\rangle$. At the crossing,
\begin{equation}
\frac{q_u}{D_u}=\frac{q_v}{D_v}\equiv s_c.
\end{equation}

Set $\bm p=\bm k-\bm k_c$, $\delta\stiffness=\stiffness(\bm k_c+\bm p)-\stiffness_c$, and $\bm G_c=(\stiffness_c-\mathrm i c_\omega\bm I_4)^{-1}$. Since $\bm Q$ is momentum independent,
\begin{equation}
\delta\bm S=\delta\bm G\bm Q\bm G_c^\dagger+ \bm G_c\bm Q\delta\bm G^\dagger+\mathcal O(\bm p^2), \qquad \delta\bm G=-\bm G_c \delta\stiffness \bm G_c+\mathcal O(\bm p^2).
\end{equation}
Its off-diagonal element in the subspace spanned by $|u\rangle$ and $|v\rangle$ is
\begin{equation}
\langle u|\delta\bm S|v\rangle =-\langle u|\delta\stiffness|v\rangle \left[ \frac{q_v}{(\lambda_u-\mathrm i c_\omega)D_v} +\frac{q_u}{D_u(\lambda_v+\mathrm i c_\omega)} \right]+\mathcal O(\bm p^2),
\label{eq:app_general_delta_S}
\end{equation}
which at the crossing becomes
\begin{equation}
\langle u|\delta\bm S|v\rangle =-s_c\langle u|\delta\stiffness|v\rangle \left[ \frac{1}{\lambda_u-\mathrm i c_\omega} +\frac{1}{\lambda_v+\mathrm i c_\omega} \right]+\mathcal O(\bm p^2).
\label{eq:app_delta_S_at_crossing}
\end{equation}
With $\bm V_c=(|u\rangle,|v\rangle)$, the reduced $2\times2$ matrix for the two relevant bands is $\bm S_{\rm eff}=\bm V_c^\dagger\bm S\bm V_c$. At each crossing below, direct expansion also gives $\langle r|\delta\stiffness|r\rangle=\mathcal O(\bm p^2)$ for $r=u,v$, so the projected diagonal elements contain no term linear in $\bm p$.

We now specialize our analysis to specific closings.

\subsection{Middle-band closings at \texorpdfstring{$\bm{K}$}{K} and \texorpdfstring{$\bm{K}'$}{K'}}
\label{app:valley-blocks}
\label{app:effective-K}
\label{app:effective-Kprime}

Using the circular vectors introduced in Eq.~\eqref{eq:circular_basis}, we define $|\alpha,\sigma\rangle=|\alpha\rangle\otimes\bm e_\sigma$ for sublattice $\alpha\in\{A,B\}$ and polarization $\sigma\in\{+,-\}$. The bond matrix obeys
\begin{equation}
\bm T(\bm{K})=-\frac32\bm e_-\bm e_+^\dagger,\qquad \bm T(\bm{K}')=-\frac32\bm e_+\bm e_-^\dagger.
\end{equation}
Thus, in the ordered basis
\begin{equation}
\mathcal B_K=(|A,+\rangle,|B,-\rangle,|A,-\rangle,|B,+\rangle),
\end{equation}
the stiffness and chiral forcing spectrum matrices at $\bm{K}$ are
\begin{equation}
\stiffness(\bm{K})= \begin{pmatrix} \lambda_A&0&0&0\\
0&\lambda_B&0&0\\
0&0&\lambda_A&t_\kappa\\
0&0&t_\kappa&\lambda_B \end{pmatrix}, \qquad \bm Q=\operatorname{diag}(q_+,q_-,q_-,q_+).
\end{equation}
The states $|A,+\rangle$ and $|B,-\rangle$ are therefore eigenvectors of $\bm S(\bm{K})$, with eigenvalues $q_+/D_A$ and $q_-/D_B$; these are the crossing bands $s_2$ and $s_3$. At $\bm{K}'$, the corresponding pair is $|A,-\rangle$ and $|B,+\rangle$. To make the Dirac velocity real and positive, define
\begin{equation}
e^{\mathrm i\theta}=\frac{A_c^*}{|A_c|},\qquad A_c=\frac{1}{\lambda_A-\mathrm i c_\omega}+\frac{1}{\lambda_B+\mathrm i c_\omega},
\end{equation}
and choose the phase-adjusted bases
\begin{equation}
\bm V_K=(|A,+\rangle,-e^{\mathrm i\theta}|B,-\rangle),\qquad \bm V_{K'}=(|A,-\rangle,e^{\mathrm i\theta}|B,+\rangle).
\end{equation}
This constant phase choice does not change the eigenstates' physical content.
For $\nu\in\{K,K'\}$, the mean and signed half-difference of the selected levels are
\begin{equation}
s_0^{(\nu)}=\frac{q_0}{2}\left(\frac{1+\eta_\nu\chi}{D_A} +\frac{1-\eta_\nu\chi}{D_B}\right), \qquad m_\nu=\frac{q_0}{2}\left(\frac{1+\eta_\nu\chi}{D_A} -\frac{1-\eta_\nu\chi}{D_B}\right),
\end{equation}
with $\eta_K=+1$, $\eta_{K'}=-1$. Thus $m_\nu=0$ at $\chi=\eta_\nu\chi_K$, where
\begin{equation}
\chi_K=\frac{D_A-D_B}{D_A+D_B},\qquad s_c^{(K)}=\frac{2q_0}{D_A+D_B}.
\end{equation}

For $\bm p=\bm k-\bm k_K$ at $\bm{K}$ and $\bm p=\bm k-\bm k_{K'}$ at $\bm{K}'$, the matrix elements between the unphased circular states are
\begin{equation}
\langle A,+|\delta\stiffness|B,-\rangle=\frac{3\kappa}{4}(p_x-\mathrm i p_y)+\mathcal O(\bm p^2), \qquad \langle A,-|\delta\stiffness|B,+\rangle=-\frac{3\kappa}{4}(p_x+\mathrm i p_y)+\mathcal O(\bm p^2).
\end{equation}
Using Eq.~\eqref{eq:app_delta_S_at_crossing} then yields
\begin{equation}
\bm S_{\mathrm{eff}}^{(\nu)} =s_0^{(\nu)}\bm I_2 +\mathv_K(p_x\bm\sigma_x+\eta_\nu p_y\bm\sigma_y) +m_\nu\bm\sigma_z +\mathcal O \left(\bm p^2+|\bm p||\chi-\eta_\nu\chi_K|\right).
\label{eq:app_Seff_K_final}
\end{equation}
Here
\begin{equation}
\mathv_K=\frac{3\kappa s_c^{(K)}(\lambda_A+\lambda_B)}{4\sqrt{D_A D_B}}.
\end{equation}
The two valley windings are opposite, $J_K=+\mathv_K^2$ and $J_{K'}=-\mathv_K^2$.

\subsection{Middle-band closing at \texorpdfstring{$\bm{\Gamma}$}{Gamma}}
\label{app:effective-Gamma}

At $\bm{\Gamma}$, the stiffness factorizes into polarization and sublattice sectors:
\begin{equation}
\stiffness(\bm{\Gamma})=\stiffness_{\rm AB}(\bm{\Gamma})\otimes\bm I_2,\qquad \stiffness_{\rm AB}(\bm{\Gamma})=\begin{pmatrix}\lambda_A&-t_\kappa\\
-t_\kappa&\lambda_B\end{pmatrix}.
\end{equation}
The sublattice matrix $\stiffness_{\rm AB}$ has eigenvalues
\begin{equation}
\Lambda_{1,2}=\frac{\lambda_A+\lambda_B}{2} \mp\sqrt{\left(\frac{\lambda_A-\lambda_B}{2}\right)^2+t_\kappa^2},
\end{equation}
with eigenvectors $|\phi_j\rangle$. The states $|\phi_j,\sigma\rangle=|\phi_j\rangle\otimes\bm e_\sigma$ are common eigenvectors of $\stiffness$ and $\bm Q$. The corresponding eigenvalues of $\bm S(\bm{\Gamma})$ are therefore
\begin{equation}
\widetilde s_{j,\sigma}(\bm{\Gamma})=\frac{q_\sigma}{L_j}, \qquad L_j=\Lambda_j^2+c_\omega^2,\qquad L_1<L_2.
\end{equation}
The two levels that cross at $\chi=\pm\chi_\Gamma$ are $s_2$ and $s_3$. The superscript $(\Gamma,\pm)$ specifies the corresponding transition, with the ordered bases
\begin{equation}
\bm V^{(\Gamma,\pm)}=(|\phi_2,\pm\rangle,|\phi_1,\mp\rangle).
\end{equation}
At $\bm p=\bm0$, the reduced matrix is $s_0^{(\Gamma,\pm)}\bm I_2+m^{(\Gamma,\pm)}\bm\sigma_z$, where
\begin{equation}
s_0^{(\Gamma,\pm)}=\frac{q_0}{2}\left(\frac{1\pm\chi}{L_2}+\frac{1\mp\chi}{L_1}\right),\qquad m^{(\Gamma,\pm)}=\frac{q_0}{2}\left(\frac{1\pm\chi}{L_2}-\frac{1\mp\chi}{L_1}\right) =\frac{q_0}{2}\left(\frac1{L_1}+\frac1{L_2}\right)(\pm\chi-\chi_\Gamma).
\end{equation}
The positive threshold magnitude and the crossing eigenvalues are
\begin{equation}
\chi_\Gamma=\frac{L_2-L_1}{L_2+L_1} = \frac{\Lambda_2^2-\Lambda_1^2}{\Lambda_2^2+\Lambda_1^2+2c_\omega^2}>0, \qquad s_c^{(\Gamma,\pm)}=\frac{2q_0}{L_1+L_2}.
\end{equation}
Thus $m^{(\Gamma,+)}=0$ at $\chi=+\chi_\Gamma$, and $m^{(\Gamma,-)}=0$ at $\chi=-\chi_\Gamma$.

The linear stiffness matrix elements are
\begin{equation}
\langle\phi_2,\pm|\delta\stiffness|\phi_1,\mp\rangle =\pm\frac{3\kappa}{4}(p_x\pm\mathrm i p_y)+\mathcal O(\bm p^2), \qquad \bm p=\bm k-\bm k_\Gamma.
\end{equation}
Using Eq.~\eqref{eq:app_delta_S_at_crossing} and choosing a constant relative phase in each basis to make the velocity real and positive gives
\begin{equation}
\bm S_{\mathrm{eff}}^{(\Gamma,\pm)}=s_0^{(\Gamma,\pm)}\bm I_2 +\mathv^{(\Gamma,\pm)}(p_x\bm\sigma_x\mp p_y\bm\sigma_y) +m^{(\Gamma,\pm)}\bm\sigma_z+\mathcal O\left(\bm p^2+|\bm p||\chi\mp\chi_\Gamma|\right).
\label{eq:app_Seff_Gamma}
\end{equation}
Here
\begin{equation}
\mathv^{(\Gamma,\pm)}=\frac{3\kappa s_c^{(\Gamma,\pm)}(\Lambda_1+\Lambda_2)}{4\sqrt{L_1L_2}}, \qquad J^{(\Gamma,\pm)}=\mp[\mathv^{(\Gamma,\pm)}]^2.
\end{equation}
The negative-polarization basis exchanges the two circular polarizations and reverses the winding. As $|\chi|$ increases through $\chi_\Gamma$, either mass changes from negative to positive, giving $\delta\mathscr C_2=\mp1$ for the $(\Gamma,\pm)$ transition.

\section{Quadratic outer-gap touchings at \texorpdfstring{$\bm{\Gamma}$}{Gamma}}
\label{app:quadratic-Gamma}

We derive the effective two-band matrices for the outer-gap closings at $\chi=0$ in Fig.~\ref{fig:all-gap-chern}. We follow the same steps as in Appendix~\ref{app:effective-correlation-theory} and still use $c_\omega=\gamma\omega$, $t_\kappa=3\kappa/2$, and $L_j=\Lambda_j^2+c_\omega^2$. We consider the pinned case, $\Lambda_1>0$.

\subsection{The touching states at \texorpdfstring{$\bm{\Gamma}$}{Gamma}}

As in Appendix~\ref{app:effective-Gamma}, the stiffness at $\bm{\Gamma}$ separates into a sublattice matrix and an identity in polarization space. Its two distinct eigenvalues are $\Lambda_1<\Lambda_2$, with sublattice eigenvectors $|\phi_j\rangle$. Each sublattice pattern has two circular polarizations, so the four states are $|\phi_j,\sigma\rangle=|\phi_j\rangle\otimes\bm e_\sigma$, with $j\in\{1,2\}$ and $\sigma\in\{+,-\}$. Their correlation eigenvalues are
\begin{equation}
\widetilde s_{j,\pm}(\bm{\Gamma})=\frac{q_0(1\pm\chi)}{L_j}.
\end{equation}
At $\chi=0$, the two polarizations of each stiffness eigenvalue are degenerate. Since $L_1<L_2$, their ordering is
\begin{equation}
s_1(\bm{\Gamma})=s_2(\bm{\Gamma})=\frac{q_0}{L_2},\qquad s_3(\bm{\Gamma})=s_4(\bm{\Gamma})=\frac{q_0}{L_1}.
\end{equation}
Thus $j=2$ describes the lower pair ($s_1$ and $s_2$) and $j=1$ the upper pair ($s_3$ and $s_4$). The middle-gap closing of Appendix~\ref{app:effective-Gamma} involves one state from each sector. We use the basis:
\begin{equation}
\bm V_j=(|\phi_j,+\rangle,|\phi_j,-\rangle).
\end{equation}
At $\bm{\Gamma}$, the reduced correlation matrix is therefore
\begin{equation}
\bm S_{\mathrm{eff},j}^{(\Gamma)}(\bm k = \bm{\Gamma})\equiv \bm V_j^\dagger \bm S(\bm{\Gamma})\bm V_j=\frac{q_0}{L_j}\bm I_2+m_j^{(\Gamma)}\bm\sigma_z,\qquad m_j^{(\Gamma)}=\frac{q_0\chi}{L_j}.
\end{equation}
Chirality splits the two polarizations by $2|m_j^{(\Gamma)}|$. To determine the nature of their touching at $\chi=0^\pm$, we now calculate how they separate slightly away from $\bm{\Gamma}$.

\subsection{Momentum dependence near \texorpdfstring{$\bm{\Gamma}$}{Gamma}}

Set $\bm p=\bm k-\bm{\Gamma}$. At $\chi=0$, the force spectrum is $q_0\bm I_4$, and Eq.~\eqref{eq:nonchiral_S} gives
\begin{equation}
\bm S=f(\stiffness),\qquad f(\lambda)={q_0}/({\lambda^2+c_\omega^2}).
\end{equation}
The stiffness and correlation matrices then share their eigenvectors. We can therefore obtain the momentum dependence of each correlation pair by first finding that of its stiffness pair.

Write a stiffness eigenvector as $(\bm u_A,\bm u_B)^{\T}$, where each sublattice amplitude has two polarization components, and denote its eigenvalue by $\lambda$. Equation~\eqref{eq:bl_stiffness} gives
\begin{equation}
(\lambda_A-\lambda)\bm u_A=\kappa\bm T\bm u_B,\qquad (\lambda_B-\lambda)\bm u_B=\kappa\bm T^\dagger\bm u_A.
\end{equation}
Solving the second equation for $\bm u_B$ and substituting into the first reduces the problem to
\begin{equation}
(\lambda_A-\lambda)(\lambda_B-\lambda)\bm u_A=\kappa^2\bm T\bm T^\dagger\bm u_A.
\label{eq:app_quadratic_elimination}
\end{equation}
Once $\bm u_A$ is known, $\bm u_B=\kappa\bm T^\dagger\bm u_A/(\lambda_B-\lambda)$  follows.

Expanding the bond matrix of Eq.~\eqref{eq:bond_matrix} around $\bm{\Gamma}$ gives
\begin{equation}
\bm T(\bm{\Gamma}+\bm p)\bm T^\dagger(\bm{\Gamma}+\bm p)=\frac94\bm I_2-\frac98\bm p\bm p^{\T}+\mathcal O(|\bm p|^3).
\end{equation}
The linear contributions from $\bm T$ and $\bm T^\dagger$ cancel, and the first term that distinguishes polarizations is quadratic: $\bm p\bm p^{\T}$.

Using Eq.~\eqref{eq:app_quadratic_elimination} yields
\begin{equation}
(\lambda-\Lambda_1)(\lambda-\Lambda_2)\bm u_A=-\frac{9\kappa^2}{8}\bm p\bm p^{\T}\bm u_A+\mathcal O(|\bm p|^3).
\end{equation}
For the pair associated with $\Lambda_j$, write $\lambda=\Lambda_j+\delta\lambda$ and retain the leading term in $\delta\lambda$. The resulting two-band stiffness matrix is
\begin{equation}
\stiffness_{\mathrm{eff},j}\simeq\Lambda_j\bm I_2+\frac{9\kappa^2}{8(\Lambda_{3-j}-\Lambda_j)}\bm p\bm p^{\T}.
\label{eq:app_quadratic_stiffness}
\end{equation}
Here $3-j$ is used to simply denote the other stiffness sector.

Applying $f$ (Eq.~\eqref{eq:nonchiral_S}) to Eq.~\eqref{eq:app_quadratic_stiffness} gives the correlation matrix to quadratic order:
\begin{equation}
\bm S_{\mathrm{eff},j}^{(\Gamma)}(\bm p)=\left({q_0}/{L_j}+\beta_j\bm p^2\right)\bm I_2+\beta_j\left[(p_x^2-p_y^2)\bm\sigma_x+2p_xp_y\bm\sigma_y\right]+m_j^{(\Gamma)}\bm\sigma_z + \mathcal O(|\bm p|^3+|\chi||\bm p|+\chi^2),
\label{eq:app_quadratic_S}
\end{equation}
with
\begin{equation}
\beta_j=\frac{9\kappa^2 f'(\Lambda_j)}{16(\Lambda_{3-j}-\Lambda_j)}=-\frac{9q_0\kappa^2\Lambda_j}{8L_j^2(\Lambda_{3-j}-\Lambda_j)},\qquad m_j^{(\Gamma)}=\frac{q_0\chi}{L_j},
\label{eq:app_quadratic_coefficients}
\end{equation}
and where we used:
\begin{equation}
\bm p\bm p^{\T}=\frac12\left[\bm p^2\bm I_2+(p_x^2-p_y^2)\bm\sigma_x+2p_xp_y\bm\sigma_y\right].
\end{equation}

\twocolumngrid
\bibliography{bib}
\end{document}